\documentclass[]{aa}  

\usepackage{graphicx}
\usepackage{txfonts}
\usepackage{upgreek}
\usepackage[]{hyperref}
\usepackage{array}
\usepackage[letterspace=150]{microtype}   

\begin{document}

\title{The Close AGN Reference Survey (CARS)}
\subtitle{Comparative analysis of the structural properties of star-forming and non-star-forming galaxy bars\thanks{Based on observations collected at the European Organisation for
Astronomical Research in the Southern Hemisphere under ESO programme(s) 094.B-0345(A) and 095.B-0015(A)}}

\titlerunning{CARS: Star formation in galaxy bars}

   \author{J. Neumann\inst{\ref{inst1}, \ref{inst2}}                            
                \and D.A. Gadotti\inst{\ref{inst3}}
                \and L. Wisotzki\inst{\ref{inst1}}
                \and B. Husemann\inst{\ref{inst4}}
                \and G. Busch\inst{\ref{inst5}}
                \and F. Combes\inst{\ref{inst6}}
                \and \\S.M. Croom\inst{\ref{inst7}}
                \and T.A. Davis\inst{\ref{inst8}}
                \and M. Gaspari\inst{\ref{inst9}}\thanks{\textit{Lyman Spitzer Jr.} Fellow}
                \and M. Krumpe\inst{\ref{inst1}}
                \and M.A. P\'erez-Torres\inst{\ref{inst10}}
                \and J. Scharw\"achter\inst{\ref{inst11}}
                \and \\I. Smirnova-Pinchukova\inst{\ref{inst4}}
                \and G.R. Tremblay\inst{\ref{inst12}} 
                \and T. Urrutia\inst{\ref{inst1}}
                }
        \institute{Leibniz-Institut f\"ur Astrophysik Potsdam (AIP), An der Sternwarte 16, D-14480 Potsdam, Germany\label{inst1}\\                                  \email{jusneuma.astro@gmail.com}
                \and European Southern Observatory (ESO), Alonso de Córdova 3107, Casilla 19001, Santiago, Chile\label{inst2}
                \and European Southern Observatory (ESO), Karl-Schwarzschild-Str.2, 85748 Garching b. München, Germany\label{inst3}
                \and Max-Planck-Institut f\"ur Astronomie, Königstuhl 17, 69117 Heidelberg, Germany\label{inst4}
                \and I. Physikalisches Institut der Universit\"at zu K\"oln, Z\"ulpicher Str. 77, 50937, K\"oln, Germany\label{inst5}
                \and Observatoire de Paris, LERMA, Coll\`ege de France, CNRS, PSL Univ., Sorbonne University, UPMC, Paris, France\label{inst6}
                \and Sydney Institute for Astronomy, School of Physics, University of Sydney, NSW 2006, Australia\label{inst7}
                \and School of Physics \& Astronomy, Cardiff University, Queens Buildings, The Parade, Cardiff, CF24 3AA, UK\label{inst8}
                \and Department of Astrophysical Sciences, Princeton University, 4 Ivy Lane, Princeton, NJ 08544-1001, USA\label{inst9}
                \and Instituto de Astrof\'isica de Andaluc\'ia, Glorieta de las Astronom\'ia, s/n, E-18008 Granada, Spain\label{inst10}
                \and Gemini Observatory, 670 N. A'ohoku Pl., Hilo, HI 96720, USA\label{inst11}
                \and Harvard-Smithsonian Center for Astrophysics, 60 Garden St., Cambridge, MA 02138, USA\label{inst12}
                }

   \date{Received 17/10/2018; accepted 29/04/2019}

\abstract{

The absence of star formation in the bar region that has been reported for some galaxies can theoretically be explained by shear. However, it is not clear how star-forming (SF) bars fit into this picture and how the dynamical state of the bar is related to other properties of the host galaxy. We used integral-field spectroscopy from VLT/MUSE to investigate how star formation within bars is connected to structural properties of the bar and the host galaxy. We derived spatially resolved H$\alpha$ fluxes from MUSE observations from the CARS survey to estimate star formation rates in the bars of 16 nearby ($0.01 < z < 0.06$) disc galaxies with stellar masses between $10^{10}\ M_\sun$ and $10^{11}\ M_\sun$. We further performed a detailed multicomponent photometric decomposition on images derived from the data cubes. We find that bars clearly divide into SF and non-star-forming (non-SF) types, of which eight are SF and eight are non-SF. Whatever the responsible quenching mechanism is, it is a quick process compared to the lifetime of the bar. The star formation of the bar appears to be linked to the flatness of the surface brightness profile in the sense that only the flattest bars $\left(n_\mathrm{bar} \leq 0.4\right)$ are actively SF $\left(\mathrm{SFR_{b}} > 0.5\ M_\sun\ \mathrm{yr^{-1}}\right)$. Both parameters are uncorrelated with Hubble type. We find that star formation is 1.75 times stronger on the leading than on the trailing edge and is radially decreasing. The conditions to host non-SF bars might be connected to the presence of inner rings. Additionally, from testing an AGN feeding scenario, we report that the star formation rate of the bar is uncorrelated with AGN bolometric luminosity. The results of this study may only apply to type-1 AGN hosts and need to be confirmed for the full population of barred galaxies.

}


\maketitle



%

\section{Introduction}
\label{Sect:Intro}

One of the main questions that is of great importance in our understanding of the formation and evolution of galaxies is which processes are responsible for quenching and triggering star formation. Bars play a major role in the redistribution of baryons and dark matter, and therefore bars are expected to have a significant effect on where and when star formation can occur.

Galactic bars are commonly observed elongated stellar structures across galaxy discs. These structures have ellipticities and lengths of varying sizes with median values of the order of $\epsilon \approx 0.6$ and $L_\mathrm{bar} \approx 4.5\,\mathrm{kpc}$, respectively \citep{Gadotti2011}. They form spontaneously from disc instabilities either in secular evolution or induced during a fly-by or merger event. The fraction of bars in disc galaxies in the local Universe is as high as $70\%$--$80\%$ \citep[e.g.][]{Eskridge2000,Menendez-Delmestre2007,Aguerri2009,Masters2011,Buta2015,Erwin2018}.
Bars are very important for many internal processes and work as engine of secular evolution and dynamics of disc galaxies \citep[e.g.][]{Kormendy2004}. These structures transfer angular momentum outwards and funnel gas to the centre of the galaxy, where it can build up structures such as nuclear rings and disc-like bulges \citep{Debattista2006,Athanassoula2013a,Sellwood2014}. Bars may feed supermassive black holes and trigger nuclear starbursts, although there is no clear correlation between the presence of a bar and an active galactic nucleus (AGN) \citep{Shlosman1989, Ho1997,Coelho2011,Cheung2015}.

The effect of stellar bars on star formation activity in the host galaxy is a widely discussed subject. It has been shown by several authors that bars are responsible for an enhancement of central star formation caused by gas inflow through the bar \citep[e.g.][]{Hawarden1986, Martinet1997, Lin2017, Catalan-Torrecilla2017}. Yet, the global star formation rate (SFR) seems not to depend on the presence of a bar \citep{Kennicutt1994a} or might even be lower for barred galaxies \citep{Cheung2013,Kim2017}. These observations are in agreement with a theory in which bars transport gas towards the centre, where it triggers star formation and the gas reservoir gets depleted, which is followed by a decrease of the global SFR. 

In this study, we shed light on another local aspect of the interplay between bars and star formation. An interesting phenomenon was observed and discussed in \citet{Garcia-Barreto1996}, \citet{Phillips1993k}, and \citet{Phillips1996}: some galaxies show a significant amount of star formation within the bar component itself, while others have bars that are quiescent. These studies also point out that late-type spiral galaxies preferably have star-forming (SF) bars, while early-type spirals have non-star-forming (non-SF) bars. \citet{Ryder1993a} found a reciprocal relationship between the number of \ion{H}{II} regions in the bar and in an inner ring component. \citet{Verley2007c} proposed three different classes of barred galaxies with respect to its star formation activity based on $\mathrm{H}\alpha$ measurements: (1) galaxies with strong central star formation, star formation at the ends of the bar and in the spiral arms, and no star formation within the bar; (2) smooth galaxies with no central star formation;  and (3) galaxies with star formation within the bar region. These authors explained these categories with an evolutionary sequence from (3) via (1) to (2), where gas is driven towards the centre and subsequently consumed. Interestingly, they could not reproduce the quiescent bars in category (1) with simulations. In these simulations, SFR calculations are mainly based on gas densities. So, how is the star formation inhibited in these bars?

In a study of CO(1-0) observations of the barred galaxy NGC1530, \citet{Reynaud1998} found strong velocity gradients in the molecular gas in the bar, which coincide with regions of weak $\mathrm{H\alpha}$. These velocity gradients perpendicular to the bar major axis cause a shearing effect on the gas. The authors argued that the shear could prevent gas clouds that are travelling along the bar to collapse and form stars. These observed velocity gradients agree with simulations in which they are associated with straight dust lanes on the leading edge of strong bars \citep[e.g.][]{Athanassoula1992}. In a sub-parsec resolution Milky Way-like simulation \citet{Emsellem2015} showed the distribution of shear, gas density, and star formation across the galaxy. In summary, these authors found that stars are forming in regions of high gas density and low shear, that is at the end of the bar and in the spiral arms, while along the central part of the bar the shear is strongest and no stars are formed. Using the same type of simulation, \citet{Renaud2015b} pointed out that the tangential velocity gradient is much smaller at the edge of the bar than in the innermost region, which makes star formation more likely to occur at the edges. Additionally, orbital crowding at the tip of the bar leads to enhanced star formation in these regions. Simulations by \citet{Khoperskov2018} have shown how the presence of a bar in massive gas-rich galaxies quenches the SFR over time both globally and within the radial extent of the bar. These authors have detected an increasing velocity dispersion within the bar region through shear during the bar formation phase that is seemingly responsible for the reduction of the SFR.

These observations and simulations provide a theory that explains the inhibition of star formation in stellar bars caused by shear. Yet, it is still not understood why some galaxies may have these velocity gradients while others may not (given that they show SF bars). If shear is the explanation of the differences seen in star formation activity in bars, then how is the presence of shear related to structural properties of the bar and the host galaxy? How is this changing during the evolutionary development of the bar?

The present work intends to contribute to a better understanding of the nature of star formation in galaxy bars by investigating major structural properties of the bars and their host galaxies in relation to the star formation activity within the bar. It makes use of spatially resolved spectroscopic data from the Multi-Unit Spectroscopic Explorer \citep[MUSE;][]{Bacon2010}, which is essential for accurately measuring emission line fluxes and being able to pinpoint the location where they were emitted. Hence, this allows us to separate star formation within the bar region from star formation outside the bar. We perform a 2D image decomposition to obtain basic parameters of the different components of the galaxy and use $\mathrm{H\alpha}$ flux from emission line fitting as tracer of star formation. We then compare the various parameters and discuss the implications of our results. Throughout the paper we assume a flat topology with a Hubble constant of $H_0 = \mathrm{67.8\,km\,s^{-1}\,Mpc^{-1}}$ and $\Omega_m=0.308$ \citep{Planck2016}.

\begin{figure*}
        \centering
        \includegraphics[width=17cm]{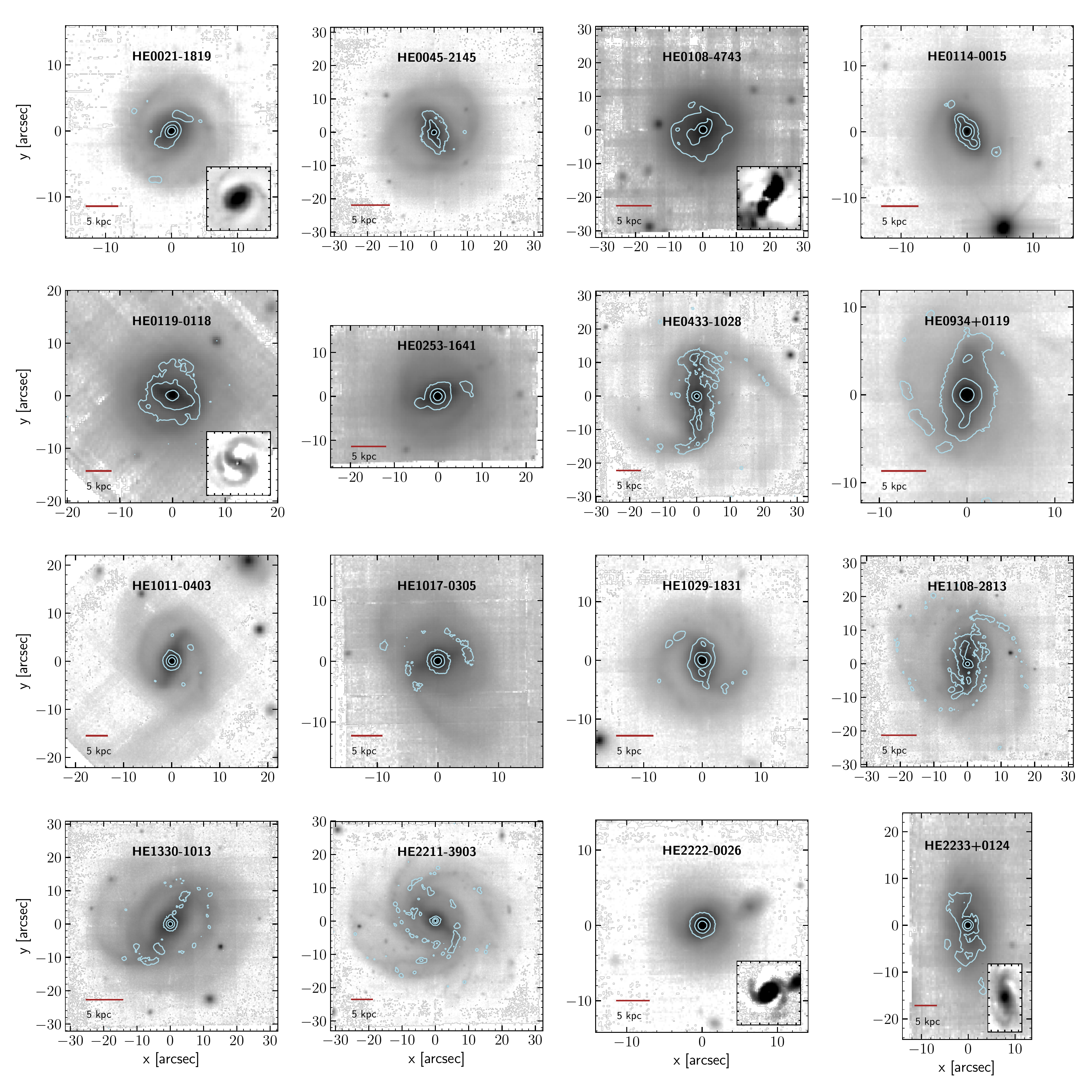}
        \caption{MUSE collapsed $i$-band images overlaid with contours of continuum subtracted $\mathrm{H\alpha}$ emission from the original data cubes (AGN+host). This figure shows our complete sample. Inset plots are added for galaxies where bars are difficult to recognise. These inset plots present residual images when subtracting a simple point source + exponential disc model from the galaxy.}
        \label{fig:all_images}
\end{figure*}

\section{Data and sample}
\label{Sect:Data}

As part of the Close AGN Reference Survey \citep[CARS;][]{Husemann2017c}\footnote{www.cars-survey.org}, this work makes use of multiwavelength observations of 41 nearby ($0.01<z<0.06$) luminous type-1 AGN host disc galaxies drawn from the Hamburg/ESO survey \citep[HES;][]{Wisotzki2000}. The analysis presented in this work is almost exclusively based on data from integral field spectroscopy (IFS) observed with MUSE on the Very Large Telescope (VLT) at Paranal. The MUSE instrument covers a 1 squared arcmin field of view (FOV) with a spatial sampling of $0.2\arcsec/\mathrm{pixel}$. It covers almost the full optical wavelength range from $470\ $nm to $930\ $nm with a mean spectral resolution of $R\sim3000$. The large FOV combined with the fine spatial sampling makes MUSE the ideal instrument to study spatially resolved spectral properties of galaxies, such as SFRs in different structural components. In addition, this study makes use of complementary infrared imaging data from SOAR/SPARTAN\footnote{The Spartan Infrared Camera \citep[SPARTAN;][]{Loh2004e} mounted on the Southern Astrophysical Research (SOAR) Telescope.} (proj. ID: 2015B-Yale/0617), NTT/SOFI\footnote{The Son OF Isaac \citep[SOFI;][]{Moorwood1998} on the New Technology Telescope (NTT).} (proj. ID: 083.B-0739(A)), and LBT/LUCI\footnote{The Large Binocular Telescope Near-infrared Utility with Camera and Integral Field Unit for Extragalactic Research \citep[LUCI;][]{Seifert2003} on the Large Binocular Telescope (LBT).} \citep[see][]{Busch2014g}, which are only included to aid in morphological galaxy classification as well as to perform sanity checks during the photometric fitting procedure.

From the CARS sample of 41 objects 37 galaxies have been observed with MUSE. We selected all galaxies that host a bar component based on our own visual classification by two of the authors, Neumann and Gadotti, in consultation with each other. Some bars may have been missed, especially very weak bars or bars in galaxies at high inclination or small apparent size. These are typical challenges and limitations that affect the studies of bars in general. However, the high quality of the data used for the classifications, namely the high signal-to-noise (S/N) images from the MUSE collapsed cubes in combination with the infrared images that are less affected by dust obscuration, favour an optimal search for bars. That selection gave us 19 barred galaxies, 2 of which had to be excluded because they did not have MUSE data and 1 galaxy was not entirely covered by the MUSE FOV and was therefore not useful for our analysis either. Our final sample comprises 16 barred galaxies of Hubble types between SBa and SBcd, with stellar masses ranging from $10^{10}$ to $10^{11}\,M_\sun$ and inclinations between approximately $0\degr < i < 63\degr$. Visual Hubble type classification was performed by two of the authors, Neumann and Gadotti, independently and then averaged. The subsample of non-barred CARS galaxies is not remarkably different from the barred galaxy sample in terms of Hubble types. Besides 9 ellipticals and 2 irregular/merger galaxies, this subsample comprises 11 galaxies of Hubble types from S0 to Sc compared to the 16 barred galaxies of types Sa to Scd. The stellar masses were estimated from ($g-i$) colours and $i$-band absolute magnitudes following the empirically calibrated relation in \citet{Taylor2011}. We used simple point-source/host-galaxy decompositions on $g$- and $i$-band collapsed MUSE images to integrate the magnitudes on the AGN-subtracted broadband images.\footnote{Emission lines were not masked when collapsing the cubes. Since the calibrated relation was established from broadband imaging of a large sample of galaxies, their contribution to the flux is already accounted for.} The inclination is estimated from the observed axial ratio of the disc component in the multicomponent decomposition described in Sect. \ref{Sect:PhotoDecomp} assuming an intrinsic thickness of $q_0=0.2$ \citep[e.g.][]{Cortese2014a}. All of the galaxies in the CARS survey were selected to host type-1 AGNs, however, owing to misclassification, 1 of our 16 galaxies (HE0045-2145) does not host an AGN. Fig. \ref{fig:all_images} shows collapsed $i$-band images from the MUSE cubes overlaid with H$\alpha$ contours of all galaxies in our sample. An overview of the main parameters of our sample can be found in Table \ref{tbl:summary_decomp}. 

The implications of the presence of an AGN on our analysis constitute an interesting and important topic. If bars are responsible for fueling AGN by driving gas inflows and if the nuclear activity depends on certain bar characteristics, then the selection of AGN host galaxies for this work could possibly introduce a bias on the type of bars and the hosts they are residing in. While we leave a full analysis with a control sample of AGN-free barred galaxies for a future paper, it is important to discuss some of the implications the selection could have on the results of this study.

We point out that our investigation of star formation along bars uses a type-1 AGN sample that avoids potential AGN misclassification depending on the Baldwin, Phillips, \& Terlevich diagram  \citep[BPT\ diagram;][see also Sect. \ref{Sect:SFR}]{Baldwin1981} selection critera for type-2 AGN as used in many previous papers studying the effect of bars on AGN fueling \citep[][]{Oh2012, Alonso2014, Galloway2015, Alonso2018}. Some BPT classifications for type-2 AGN are prone to be contaminated by low-ionisation nuclear emission-line region (LINER) galaxies which do not necessarily host AGN \citep[e.g.][]{Singh2013}. In addition, the CARS sample was drawn from the Hamburg/ESO survey selecting the most luminous AGN within a certain redshift range without applying any specific criteria on the host galaxy properties. Therefore, it is hard to study trends as a function of AGN luminosity, and the results from this study may only apply to type-1 AGN hosts. Further work is needed to confirm whether these results can be extended to the full population of barred galaxies.

The selection of barred galaxies from this sample did not introduce any obvious additional bias regarding Hubble types or galaxy properties. The presence of an AGN implies that there must be some gas in the centre of the galaxy that somehow must have been pushed inwards. However, different temporal and spatial scales of the activity of the nucleus and a large-scale bar do not permit a direct conclusion about a correlation that extends to kiloparsec scales. We elaborate on this discussion further and test an AGN feeding scenario in Sect. \ref{Sect:AGNfeeding} and briefly address AGN feedback in Sect. \ref{Sect:Conclusions}. A deeper study of AGN feeding and feedback in CARS galaxies will be the subject of upcoming papers from the collaboration.

\section{Photometric decomposition}
\label{Sect:PhotoDecomp}

Two-dimensional image decomposition has become a widely used technique to retrieve structural properties of galaxies. The accuracy, but also the degeneracy of the results depend on many factors such as the amount of detail that is desired to model, the quality of the observations, and the human based decision on which galaxy components to include in the modelling. The most basic approach is to fit a two-component bulge-disc model, which is simple enough to be conducted in an automatic way for a large sample of objects and -- to some extent -- good enough to get rough estimates of parameters such as disc scale length, bulge-to-total light ratio (B/T), or bulge S\'ersic index ($n_\mathrm{b}$) \citep[e.g.][]{Allen2006a}. However, it becomes rapidly more complex, if a higher accuracy of these parameters is desired. For example, the neglect of bars, when fitting barred galaxies, can lead to an overestimation of B/T by a factor of 2 (\citealp{Gadotti2008}; \citealp[see also][]{Aguerri2005b,Salo2015}). Similarly, not considering a point source in an AGN host galaxy, results on average in larger B/T and $n_\mathrm{b}$ and smaller effective radii of the bulge $r_\mathrm{e,b}$ \citep{Gadotti2008}. This bias is strongest in bright type-1 AGN. Moreover, the majority of galaxies have been found to have disc breaks \citep{Erwin2005a,Pohlen2006,Erwin2008,Marino2016f}. Ignoring the disc break in the fit can lead to an underestimation of B/T and bar-to-total light ratio (Bar/T) by $\sim 10\%$ and $25\%$, respectively, and differences in the disc scale length ($h$) of $\sim 40\%$ (\citealp{Kim2014a}; \citealp[see also][]{Gao2017}). Hence, it is indispensable to analyse carefully which components to include in the model.

In recent decades many programs have been made publicly available to perform 2D photometric decomposition, such as \textls{\sc gim2d} \citep{Simard2002}, \textls{\sc galfit} \citep{Peng2002,Peng2010b}, \textls{\sc budda} \citep{Souza2004}, and \textls{\sc gasp2d} \citep{Mendez-Abreu2008}. For our analysis, we used \textls{\sc imfit} (v.1.5) by \citet{Erwin2015a}\footnote{http://www.mpe.mpg.de/$\sim$erwin/code/imfit/}. The most important component for this work is the bar of the galaxy, for which we are not only interested in the basic parameters, as for example S\'ersic index ($n_\mathrm{bar}$), ellipticity ($\epsilon_\mathrm{bar}$), and Bar/T, but we also want to determine the exact region from the position, length, ellipticity, and position angle (PA) that is covered by the bar. This is important information to distinguish between the star formation that is happening within the bar and outside the bar. Bars have been modelled in 2D decompositions either with S\'ersic \citep{Sersic1963} or Ferrer \citep{Ferrers1877,Binney1987} functions \citep[e.g.][]{Laurikainen2005,Laurikainen2007,Laurikainen2010,Gadotti2008,Weinzirl2009,Peng2010b,Mendez-Abreu2017}. \citet{Kim2015} and \citet{Gao2017} showed that both profiles can describe the main shape of the bar and the results should not be influenced by the choice of the fitting function. In our model we call this component our main bar component (bar1).

\citet{Kim2015} also stressed that bars should ideally not be modelled as a single component. It is known that during the evolution of the bar it experiences a buckling instability phase that leads to a vertical thickening of the inner part of the bar with respect to the equatorial plane, as shown for example in \citet{Combes1990}, \citet{Kuijken1995a}, \citet{Athanassoula1999}, \citet{Bureau1999}, \citet{Bureau2005}, and \citet{Athanassoula2005}. In edge-on galaxies, this has been observed as boxy-, peanut-, or x-shaped feature \citep{Jarvis1986,Souza1987} and it is even visible in only moderately inclined galaxies \citep{Athanassoula2006,Erwin2013}. It appears that the same physical component manifests itself morphologically also as a barlens when seen face-on \citep{Laurikainen2011,Athanassoula2015,Laurikainen2017}. We thus adopted a two-component bar model for all the galaxies where an inner boxy bar or barlens is clearly visible; in our model, this is denoted as bar2.

\begin{figure*}
        \centering
        \includegraphics[width=17cm]{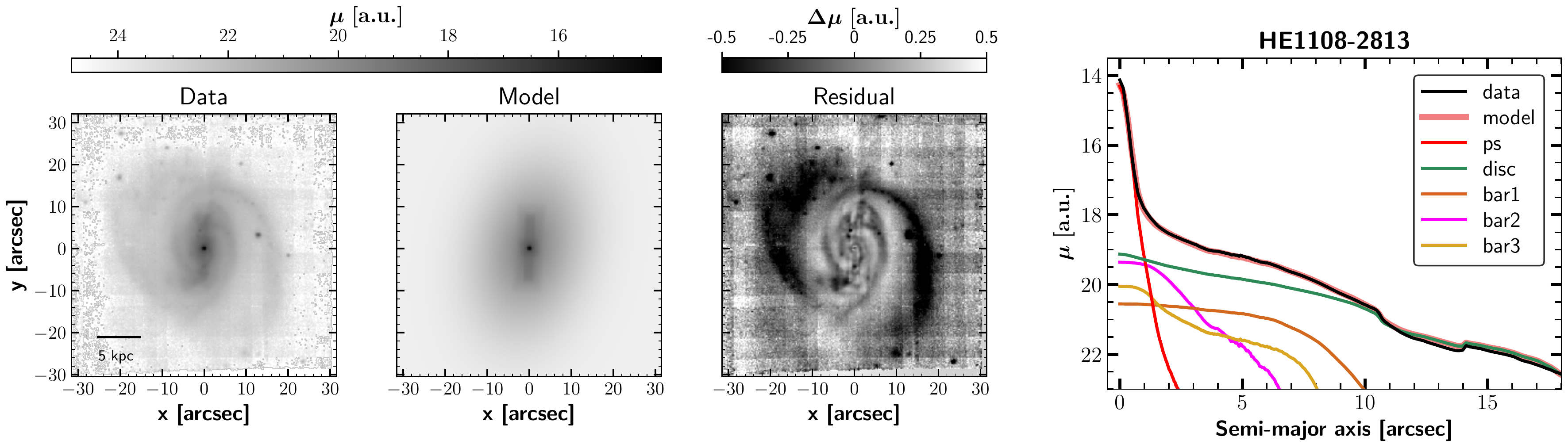}
        \caption{Photometric decomposition of MUSE collapsed $i$-band image of galaxy HE 1108-28138. From \textit{left} to \textit{right}: data image, model, residual=data-model, and surface brightness profile from isophotal fitting. The colour map in the residual image is stretched to show faint details. The surface brightness profiles in the right-most panel show a separation into all components that were included in the fit. The thick coral line shows the sum of all model components and the black line the observed data. A set of figures that shows the decomposition of the complete sample can be found in the Appendix \ref{apx:2ddecomp}.}
        \label{fig:2ddecomp}
\end{figure*}

Furthermore, in many cases there is evidence for a very elongated light excess in the residual image of the bar after the fit. It has approximately the length of the outer part of the bar, but is much thinner. It can be seen in Fig. \ref{fig:2ddecomp} and in residual images of previous works \citep[e.g.][]{Gadotti2008,Athanassoula2015}. For these galaxies, we decided to add a third bar component in the model. This decision is purely empirically motivated. We stress these are all parts of the same bar. From 2D and 3D orbital theory we know that bars are built from families of periodic orbits with different extents, elongations, and orientations. \citep[e.g.][]{Contopoulos1980,Athanassoula1983,Pfenniger1984,Skokos2002,Skokos2002a}. While the usual single bar component fits the main orbits that constitute the backbone of the bar, the very narrow extra component in the residual might be a signature of very elongated orbits along the bar major axis. We include this component in our model when necessary; we call this in our model bar3. Additionally, a ring was fitted when present. This is important to accurately determine the S\'ersic index of the main bar component.

In summary, we fit a selection of the galaxy components given in Table \ref{tbl:photo_comps} and we only include a component if we have clear visual indications of its presence. Out of the 16 galaxies 8 were modelled with a single bar component, 5 needed two components, and 3 galaxies were fitted with all three bar components.

\begin{table}
\caption{Overview of model components used in the photometric decomposition.}             
\label{tbl:photo_comps}      
\centering                          
\begin{tabular}{c c}       
\hline\hline                 
Model component & Function \\   
\hline                        
Point source & Gaussian with a 0.1 px width\\
Bulge & S\'ersic\\
Disc & Exponential\\
Bar1 & S\'ersic with generalised ellipses\\
Bar2 & S\'ersic with generalised ellipses\\
Bar3 & S\'ersic with generalised ellipses\\
Ring & Gaussian ring\\
Background & FlatSky \\
\hline                                   
\end{tabular}
\tablefoot{The three bar components given are all parts of the same bar. In detail, bar1 is the main long part of the bar, bar2 is the broadened inner part of the bar, and bar3 is the very thin and long part (see text for details).}
\end{table}

All fits were performed independently on pseudo SDSS $i$-band images from the MUSE data cubes and on $R$-band or $J$-band images from either SOAR/SPARTAN, NTT/SOFI, or LBT/LUCI. During the fitting of the $R$-band and $J$-band images, we identified some problems to model accurately the point-spread function and some of the images were taken with rather short exposure times. Furthermore, only observations from MUSE were available for the complete sample, which gives us the advantage of consistency. After a careful comparison, we opted to use only the results performed on the MUSE $i$-band images. However, the decompositions on the broadband images were a useful control set to tune the initial parameter values and evaluate intermediate results during the fitting procedure. The point-spread function was determined by fitting Moffat profiles to at least two point sources in the FOV of each image. The stripes in the background of the images as seen for example in Fig. \ref{fig:all_images} -- a known effect from the integral field unit -- are very shallow and do not affect the decomposition.

Fig. \ref{fig:2ddecomp} shows for the galaxy HE 1108-2813 from left to right the collapsed $i$-band image from MUSE, the model, residual image, and surface brightness profile. The profile was derived by fitting ellipses to the isophotes in the images with the \textls{\sc iraf}\footnote{\textls{\sc iraf} is distributed by the National Optical Astronomy Observatories, which are operated by the Association of Universities for Research in Astronomy, Inc., under cooperative agreement with the National Science Foundation.} task \textls{\sc ellipse}. In a first step, we fitted the data image with the parameters for PA and ellipticity left free. Leaving these parameters free has the advantage that the surface brightness profile highlights at each distance the predominant source that contributes to the total surface brightness. Then, we used the same set of values from the first fit to perform the same task on the model images in non-fitting mode just measuring the surface brightness on the same ellipses. This can be done via the \textls{\sc inellip} option.  A summary of the parameters from the decomposition is given in Table \ref{tbl:summary_decomp}.

\begin{table*}
\caption{Summary of the main parameters from the 2D decomposition, bar length measurement, and morphological classification.}
\label{tbl:summary_decomp}
\centering

\begin{tabular}{c c c c c c l c c c c c c c}
\hline\hline
Galaxy & Type & Incl[$\degr$] & PS/T & B/T & \multicolumn{2}{c}{Bar/T} & R/T & D/T & $h\,['']$ & $n_\mathrm{bar1}$ & $n_\mathrm{bar2}$ & $n_\mathrm{bar3}$ & $L_\mathrm{bar}\,['']$\\
(1) & (2) & (3) & (4) & (5) & \multicolumn{2}{c}{(6)} & (7) & (8) & (9) & (10) & (11) & (12) & (13)\\
\hline
\object{HE0021-1819}      &       SBab    &       16     &        0.03    &       0.20     &       0.08 &(100/0/0)         &       0.10    &       0.59    &        3.44    &       0.76    &       --      &       --      &       2.84 \\
\object{HE0045-2145}      &       SBab    &       5      &        0.16    &       0.02     &       0.23 &(68/0/32)         &       --      &       0.59    &        6.89    &       0.34    &       --      &       0.16  &         8.30 \\
\object{HE0108-4743}      &       SBc     &       13     &        0.10    &       --           &   0.10 &(100/0/0)         &       --      &       0.80    &        4.02    &       0.97    &       --      &       --      &       4.98 \\
\object{HE0114-0015}      &       SBab    &       40     &        0.05    &       0.14     &       0.27 &(75/0/25)         &       --      &       0.54    &        4.23    &       0.28    &       --      &       0.06    &       3.60 \\
\object{HE0119-0118}      &       SBab    &       2      &        0.23    &       --           &   0.12 &(100/0/0)         &       --      &       0.65    &        3.37    &       0.37    &       --      &       --      &       4.84 \\
\object{HE0253-1641}      &       SBab    &       30     &        0.39    &       --           &   0.16 &(100/0/0)         &       --      &       0.45    &        5.80    &       0.75    &       --      &       --      &       9.16 \\
\object{HE0433-1028}      &       SBcd    &       57     &        0.19    &       0.06     &       0.21 &(77/0/23)         &       --      &       0.54    &        6.00    &       0.09    &       --      &       0.05    &       14.78 \\
\object{HE0934+0119}      &       SBab    &       38     &        0.45    &       --           &   0.18 &(73/0/27)         &       --      &       0.36    &        4.98    &       0.22    &       --      &       0.05    &       6.78 \\
\object{HE1011-0403}      &       SBb     &       26     &        0.42    &       0.02     &       0.17 &(100/0/0)         &       --      &       0.39    &        4.51    &       0.50    &       --      &       --      &       6.59 \\
\object{HE1017-0305}      &       SBc     &       54     &        0.24    &       0.02     &       0.16 &(65/9/26)         &       0.01    &       0.58    &        6.83    &       0.38    &       0.05    &       0.18    &       5.37 \\
\object{HE1029-1831}      &       SBab    &       38     &        0.15    &       0.26     &       0.15 &(100/0/0)         &       --      &       0.44    &        4.45    &       0.21    &       --      &       --      &       3.78 \\
\object{HE1108-2813}      &       SBc     &       51     &        0.17    &       --           &   0.22 &(46/23/31)        &       --      &       0.62    &        5.38    &       0.26    &       0.24    &       0.07    &       9.35 \\
\object{HE1330-1013}      &       SBc     &       40     &        0.06    &       0.02     &       0.18 &(82/18/0)         &       0.06    &       0.69    &        11.20   &       0.64    &       0.26    &       --      &       11.56 \\
\object{HE2211-3903}      &       SBbc    &       4      &        0.20    &       0.02     &       0.17 &(52/36/12)        &       0.01    &       0.60    &        7.12    &       0.60    &       0.39    &       0.05    &       8.44 \\
\object{HE2222-0026}      &       SBa     &       7      &        0.42    &       --           &   0.15 &(100/0/0)         &       --      &       0.44    &        2.06    &       0.55    &       --      &       --      &       3.25 \\
\object{HE2233+0124}      &       SBb     &       63     &        0.13    &       0.11     &       0.14 &(100/0/0)         &       --      &       0.62    &        5.22    &       0.73    &       --      &       --      &       4.52 \\
\hline
\end{tabular}
\tablefoot{
(1) Galaxy name; (2) Hubble type from our own visual classification by two of the authors, Neumann and Gadotti; (3) inclination of the galaxy calculated from the ellipticity of the disc component in the decomposition assuming an intrinsic thickness of the disc of $q_0=0.2$; (4)-(8) luminosity fractions of point source, bulge, bar, ring and disc, respectively; in (6) we are additionally showing the relative contribution in percentage of each bar component (bar1/bar2/bar3) to the total luminosity of the bar; (9) disc scale length; (10)-(12) S\'ersic indices of the main, second (broadened) and third (narrow) bar component, respectively; (13) length of the bar.
}
\end{table*}

\section{Measuring the length of the bar}
\label{Sect:barlength}

There is no unambiguous way to determine the length of the bar, and authors have
determined this value in many different ways in the past; there is no sharp transition, but bars join smoothly the outer disc.  A common approach is to use the isophotal ellipticity profile of the galaxy and define the bar length as the distance from the centre (on the x-axis) at the first maximum ellipticity \citep[on the y-axis; see][]{Wozniak1991}. We call this $L_\mathrm{peak}$. This maximum is usually associated with the ellipses close to the end of the bar, after which the bar transitions into the disc causing the ellipticity to drop. If the bar is strong, this drop may be very fast, but weak bars tend to show a slow decline in ellipticity \citep[e.g.][]{Gadotti2007}.

Many different approaches of measuring the length of the bar for simulated galaxies were tested and compared in \citet{Athanassoula2002}. These authors found that the maximum ellipticity method generally provides the smallest value for the bar length. Alternatively, the first minimum after the ellipticity peak ($L_\mathrm{min}$) or a sudden change of the PA ($L_\mathrm{PA}$) can be used \citep{Erwin2003b}. \citet{Erwin2005c} showed that $L = \mathrm{minimum(}L_\mathrm{min},L_\mathrm{PA}$) correlates very well with $L_\mathrm{peak}$ with a Spearman correlation coefficient of $\rho=0.96$ and the average of both values matches best the visual bar size measurement ($L_\mathrm{vis}$).

In this work, we derived an estimate by a combination of different proxies for the bar length, such as ellipticity peak and minimum, the change in PA and -- in specific cases -- the radius of an inner ring, as explained below. We performed the fitting of the isophotes with the \textls{\sc iraf} task \textls{\sc ellipse} as described in the previous section. We then determined, where it was possible, the location of the ellipticity peak $L_\mathrm{peak}$ and the location of the proximate minimum after the drop $L_\mathrm{min}$. We further used the radial profile of the  PA to identify sudden changes in the PA after the ellipticity peak. The position where the PA changed about $10\degr$ as compared to the PA at $L_\mathrm{peak}$ is denoted as $L_\mathrm{PA}$. Our first estimate of the bar length is defined as follows:

\begin{equation}
L_\mathrm{bar,0} = \mathrm{AVG(}L_\mathrm{peak}\mathrm{,MIN(}L_\mathrm{min},L_\mathrm{PA}\mathrm{))}
.\end{equation}

It is not always possible to determine the length of the bar with the ellipticity or the PA profile. Sometimes the end of the bar transitions smoothly into a ring or spiral arms of the galaxy and no clear ellipticity drop or change in PA can be identified. In other cases the ellipticity can get distorted in the presence of strong dust lanes. Hence, a careful case-by-case evaluation including a visual inspection of the image is necessary. The final bar estimate is

\begin{equation}
L_\mathrm{bar} = \mathrm{MIN(}L_\mathrm{bar,0},L_\mathrm{ring}\mathrm{)} \quad \mathrm{or} \quad L_\mathrm{vis}
.\end{equation}

\begin{figure}
        \resizebox{\hsize}{!}{\includegraphics{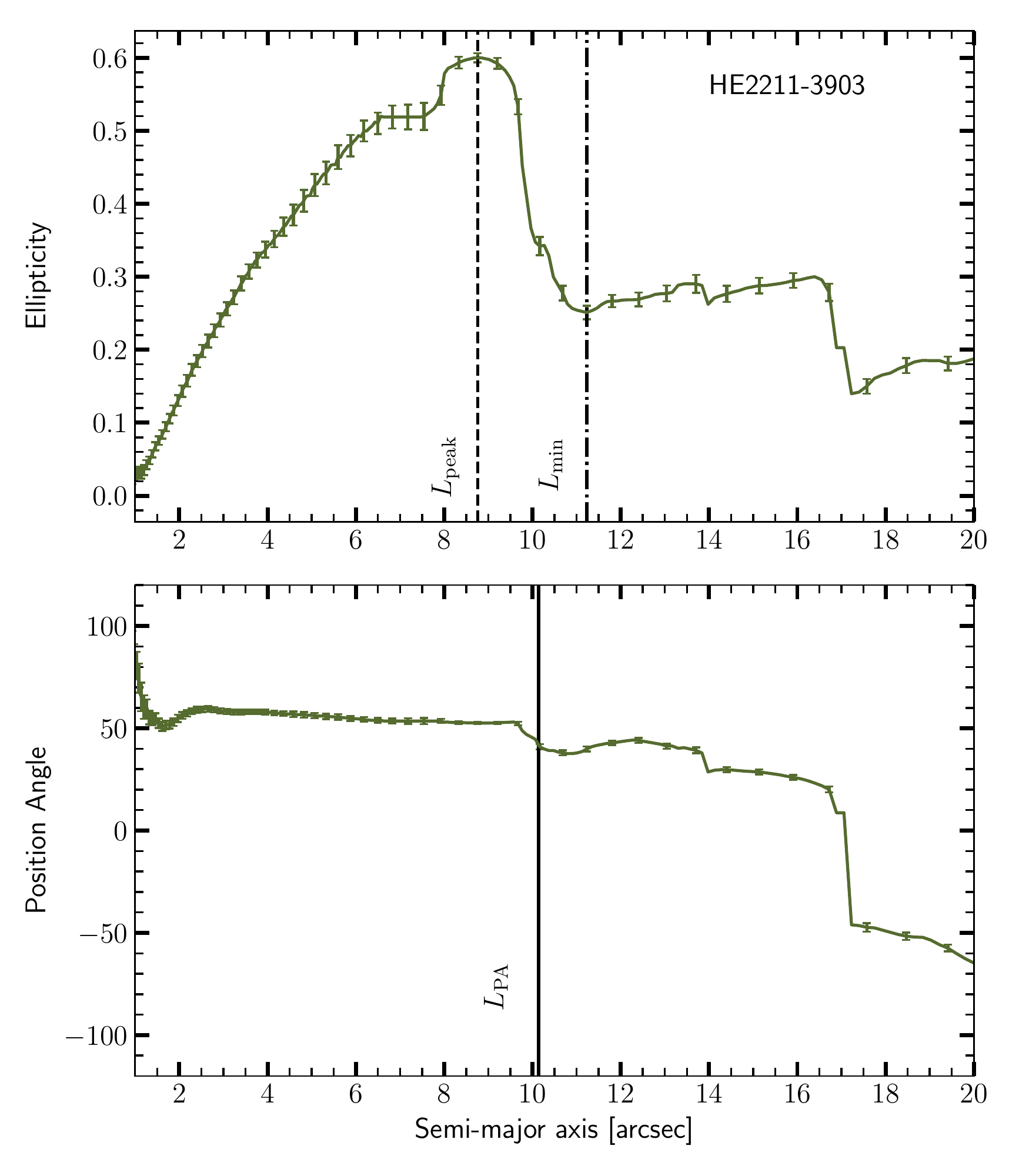}}
        \caption{Results from the isophotal fitting routine \textls{\sc ellipse} in \textls{\sc iraf} for the MUSE collapsed $i$-band image for galaxy HE 2211-3903 to estimate the bar length. The vertical lines show the different estimates of the length of the bar as described in more detail in the text.}
        \label{fig:ellipticity}
\end{figure}

We only had to use a visual estimation in the case of HE2233+0124. A typical ellipticity profile of a galaxy is exemplarily shown for HE2211-3903 in Fig. \ref{fig:ellipticity}

\section{Derivation of SFR from dust-corrected H${\alpha}$ emission lines}
\label{Sect:SFR}

Since individual stars are unresolved in our galaxy sample, measurements of star formation activity rely on tracers of star formation in the spectrum of integrated light. A big pool of diagnostic methods across the electromagnetic spectrum from ultraviolet (UV) to far-infrared has been established over the years \citep[for a review, see][]{Kennicutt1998,Kennicutt2012}. While the young stellar population can be directly observed in UV emission, this method has the disadvantage that a significant fraction of the UV light is absorbed by dust and re-emitted in the far-infrared. Furthermore, for the analysis of nearby galaxies this method is limited to space telescope observations. Alternatively, recombination lines in the optical that trace the re-emission of ionised hydrogen in \ion{H}{II} regions can be used. The $\mathrm{H\alpha}$ line has become very popular for a measurement of the SFR. This is one of the strongest emission lines and it is less affected by dust obscuration than UV tracers. For local galaxies, $\mathrm{H\alpha}$ is within the MUSE spectral range (470-930 nm). In this section, we describe the procedure for getting spatially resolved SFR maps from the MUSE observations.

Prior to the emission line analysis, a deblending of AGN and host galaxy is performed using the software \textls{\sc QDeblend3D} \citep{Husemann2013a,Husemann2014}. The stellar continuum and emission lines are then modelled on the AGN-substracted cubes with the code \textls{\sc PyParadise} \citep {Walcher2015,Husemann2016b} that uses a linear combination of stellar template spectra convolved with a Gaussian line-of-sight velocity kernel. The fitting is done in three steps. First, the continuum is fitted for co-added spectra in Voronoi bins with target S/N$\sim$10. Second, the continuum is fitted on a spaxel-by-spaxel basis with fixed kinematics according to the underlying Voronoi cell. Finally, the emission lines are modelled in the residual spectra using a single Gaussian component. A Monte Carlo simulation is used to estimate errors by refitting the data 100 times after modulating the input data by the formal errors \citep[for more details of the functionality of \textls{\sc PyParadise,} see also][]{DeRosa2018,Weaver2018}.

The $\mathrm{H\alpha}$ flux has to be corrected for dust attenuation to derive accurate SFRs. Since the intrinsic ratio of the emission lines $\mathrm{H\alpha}/\mathrm{H\beta}$ (Balmer decrement) is in ideal conditions set by quantum mechanics, it has been commonly used as a measure of the effect of dust on the source spectrum of interest. The attenuation is wavelength dependent and thus changes the observed ratio. Following \citet{Calzetti1994} and \citet{Dominguez2013} the intrinsic luminosity $L_\mathrm{int}$ can be obtained by

\begin{align}
L_{\mathrm{int}}(\lambda) = L_{\mathrm{obs}}(\lambda)\ 10^{0.4\ k(\lambda)\ E(B-V)}.
\end{align}

In this equation, $L_\mathrm{obs}$ is the observed luminosity and $k(\lambda)$ is the reddening curve. In this work we use the reddening curve from \citet{Calzetti2000}. $E(B-V)$ is the colour excess that is given by

\begin{equation}
E(B-V) = 1.97\ \log_{10} \left( \frac{(\mathrm{H\alpha}/\mathrm{H\beta})_\mathrm{obs}}{2.86}\right).
\end{equation}

After this analysis, it is possible to compute a spatially resolved dust-corrected $\mathrm{H\alpha}$ map for each galaxy. We clean this map by considering only spaxels, where $\mathrm{S/N}_{\mathrm{H\alpha}}>3$, $\mathrm{S/N}_{\mathrm{H\beta}}>3$, $V_\mathrm{err}<20\, \mathrm{km/s}$, and $\sigma_\mathrm{err}<30\, \mathrm{km/s}$.

The presence of an AGN can contaminate the measurement of $\mathrm{H\alpha}$-based SFRs. It is no longer possible to convert $\mathrm{H\alpha}$ flux to SFRs under the assumption that all $\mathrm{H\alpha}$ is caused by star formation. In contrast, a new source of photoionisation has to be considered. The AGN and star formation ionisation can be seen on emission line diagnostic diagrams, such as the traditional BPT diagram \citep[][]{Baldwin1981,Veilleux1987}. While the BPT diagram has been widely used to classify galaxies as whole systems \citep[e.g.][]{Kauffmann2003}, more recently this diagram has also been applied to analyse different regions of the galaxies with data from IFS surveys \citep{Singh2013,Belfiore2016,Federrath2017a}.

The main idea behind our analysis is to not only classify each pixel to one or the other ionisation source because the sources can be mixed and pixels form the so-called mixing sequence on a BPT diagram, but to define a fraction of H$\alpha$ per pixel caused by one mechanism or the other. We performed an emission line diagnostic that is based on an analysis described in \citet{Davies2016b} with modifications necessary for analysing more complicated BPT diagrams. We defined a number of AGN and SF basis pixels based on their spatial distribution and BPT position. To fit the mixed pixels and disentangle the AGN fraction we treated the pixels as vectors of emission line fluxes and fitted a linear combination using a Markov chain Monte Carlo (MCMC) algorithm as follows:
\begin{equation}
\begin{pmatrix}
\mbox{H}\alpha \\
\mbox{H}\beta \\
[\ion{O}{III}] \\
[\ion{N}{II}] \\
[\ion{S}{II}]
\end{pmatrix}_{\mathrm{mixed}}
=
f_{\mathrm{SF}} \times 
\begin{pmatrix}
\mbox{H}\alpha \\
\mbox{H}\beta \\
[\ion{O}{III}] \\
[\ion{N}{II}] \\
[\ion{S}{II}]
\end{pmatrix}_{\mathrm{SF}}
+
f_{\mathrm{AGN}} \times 
\begin{pmatrix}
\mbox{H}\alpha \\
\mbox{H}\beta \\
[\ion{O}{III}] \\
[\ion{N}{II}] \\
[\ion{S}{II}]
\end{pmatrix}_{\mathrm{AGN}},
\end{equation}
where $f_{\mathrm{SF}}$ and $f_{\mathrm{AGN}}$ are SF and AGN fractions and these values obey the assumption of no other excitation mechanisms $f_{\mathrm{SF}} + f_{\mathrm{AGN}} = 1$. In the fitting procedure we varied the fractions and the basis vectors, parameterised with metallicity of SF basis and the [\ion{N}{II}]/H$\alpha$ ratio of AGN basis. 

\begin{figure*}
        \centering
        \includegraphics[width=17cm]{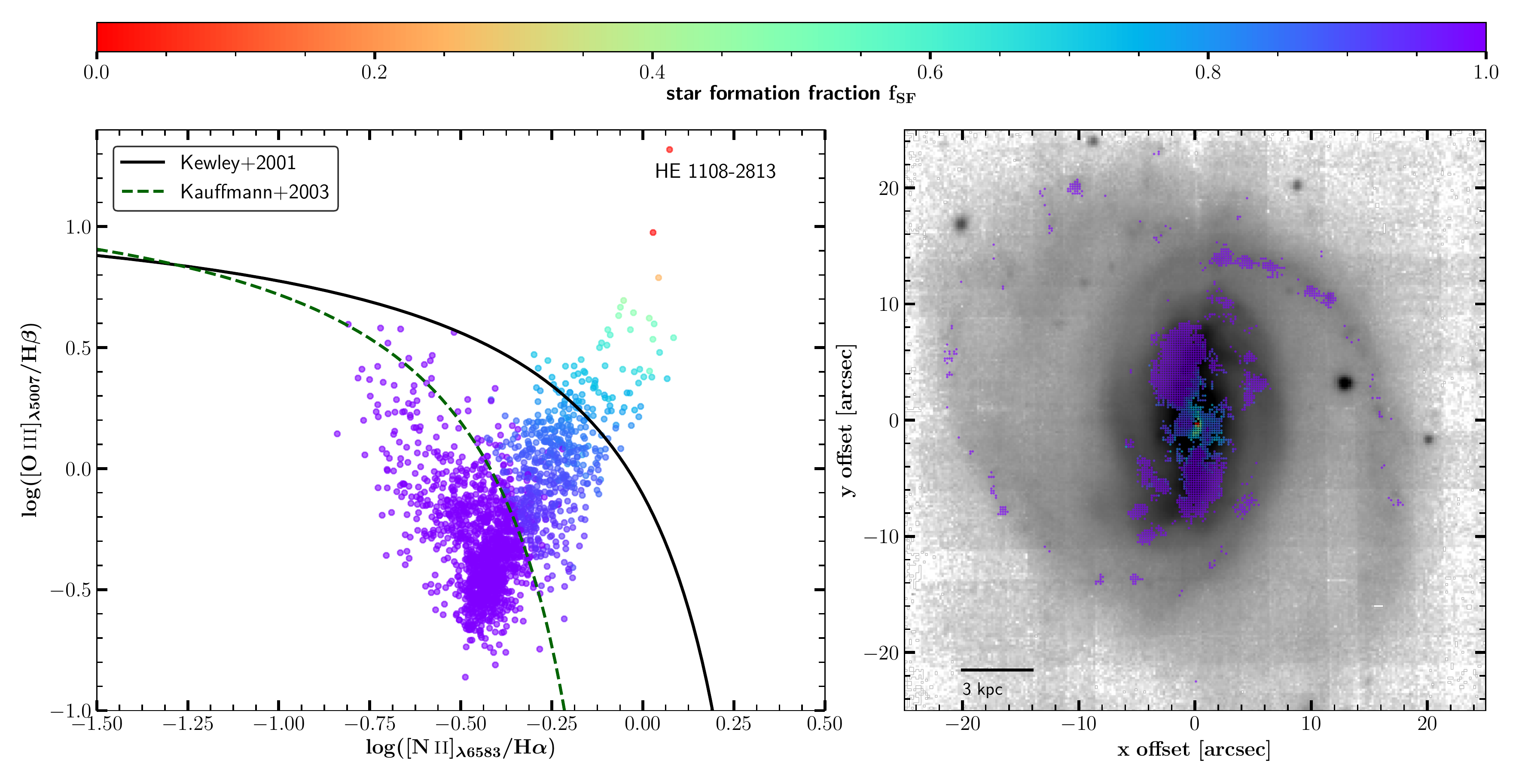}
        \caption{\textit{Left panel}: Emission line diagnostic for galaxy HE 1108-2813. Data points are coloured by the fraction of H$\alpha$ that comes from star formation. The black solid line is adopted from the theoretical line from \citet{Kewley2001} and the green dashed line from the empirically derived separation from \citet{Kauffmann2003}. We show only points where $\mathrm{S/N}>3$ for all four emission lines. \textit{Right panel}: Collapsed MUSE $i$-band image overplotted with the data points from the left panel to show the spatial location of these spectra.}
        \label{fig:BPT}
\end{figure*}

We show in Fig. \ref{fig:BPT} the spatially resolved emission line diagnostic for the galaxy HE 1108-2813. Each point in the left panel of this figure corresponds to one spectrum in the MUSE cube. The right panel shows the spatial location of the data points from the BPT diagram overplotted on the $i$-band image of the galaxy. Regions are coloured according to the fraction of H$\alpha$ that comes from star formation. For comparison, we also show the theoretical-based maximum starburst line from \citet{Kewley2001} and the empirically motivated division line from \citet{Kauffmann2003}. We note that [\ion{S}{II}] is included in the fitting, but not shown on this BPT. The figure shows that all spaxels that are most affected by the AGN are centrally concentrated and the H$\alpha$ from the bar region and the spiral arms is almost exclusively caused by star formation.

For each spaxel, we multiplied the dust-corrected H$\alpha$ flux by the star formation fraction and converted it into SFR assuming a Salpeter initial mass function \citep{Salpeter1955} and using the \citet{Kennicutt1998} relation

\begin{equation}
\mathrm{SFR}\, [M_{\sun}\, \mathrm{yr^{-1}}] = 7.9 \times 10^{-42}\, L(\mathrm{H\alpha})\, [\mathrm{erg\,s^{-1}}] \times f_\mathrm{SF}.
\end{equation}

\begin{figure}
        \resizebox{\hsize}{!}{\includegraphics{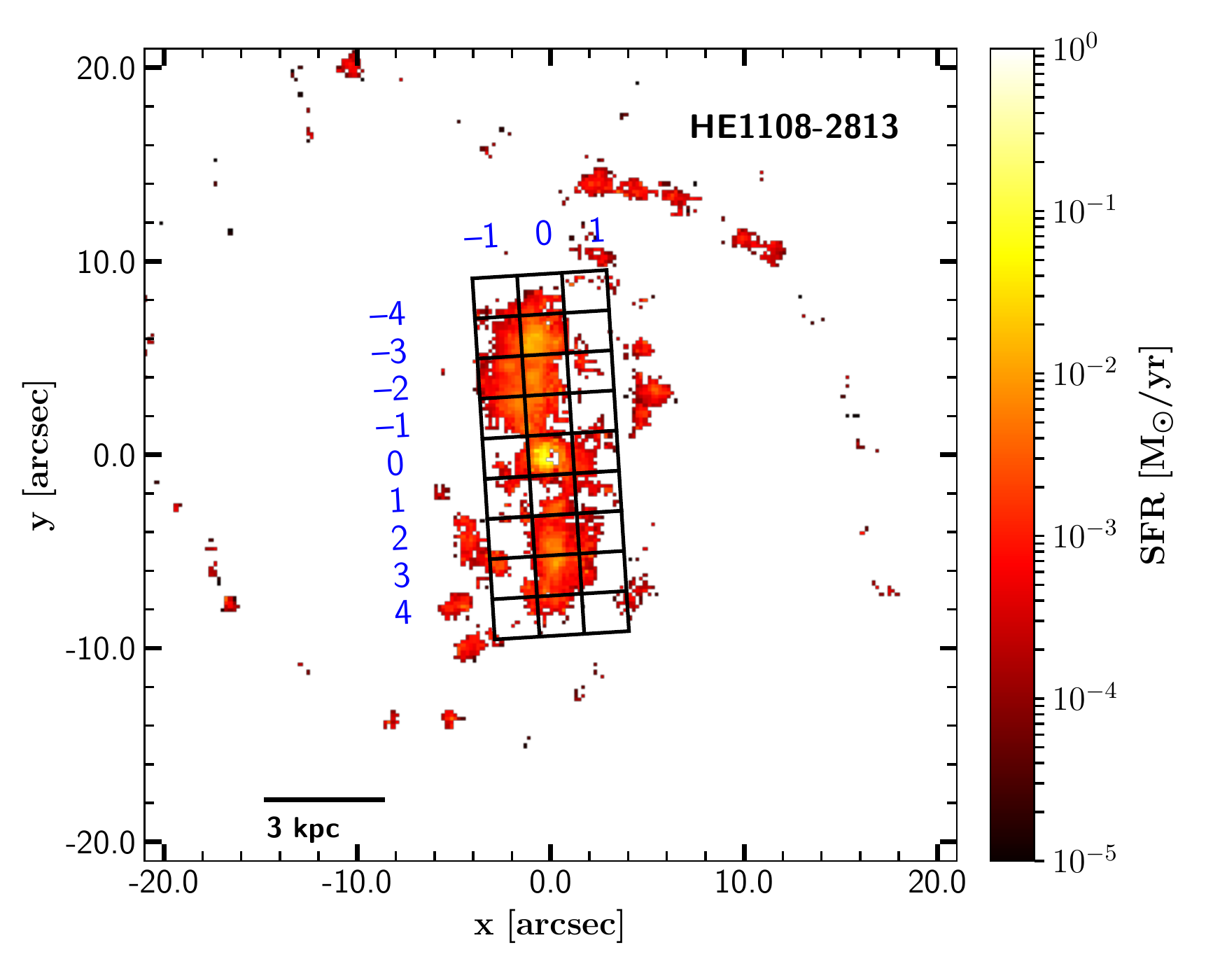}}
        \caption{Spatially resolved map of SFRs in the galaxy HE1108-2813. On top, we show the bar mask for that galaxy, created from the parameters of the image decomposition and ellipse fitting. The mask is divided into $3\times9$ subcells to analyse how the SFR changes over the bar region. A set of these maps for the complete sample with additional $i$-band contours can be found in Appendix \ref{apx:sfr}.}
        \label{fig:SFR_map}
\end{figure}

The final SFR map for the galaxy HE 1108-2813 overplotted with the bar region is shown in Fig. \ref{fig:SFR_map}. The extent of the bar is defined by the parameters of the photometric decomposition and the measurement of the bar length. As approximation, we used a rectangle with the two sides \textit{a} and \textit{b} having the length of the major and minor axis of the bar. We split the bar further into $3\times9$ subregions. A collection of these plots for the entire sample can be found in Appendix \ref{apx:sfr}.

Finally, to estimate the total SFR in the bar region ($\mathrm{SFR_{b}}$), we rebinned the original spectra in the AGN-subtracted MUSE cubes within each of the 27 bar subregions and repeated subsequently the complete analysis as described above. The decision was made to increase the S/N and get more accurate measurements of the SFR in the bar, especially in regions of low signal. Additionally, we divided the SFR by the stellar mass to derive specific star formation rates (sSFR). Throughout the analysis we used two different measures of the sSFR:

\begin{equation}
\begin{aligned}
\mathrm{s_{t}SFR_{b}} &= \mathrm{SFR_{b}}/ M_\mathrm{t}, \\
\mathrm{s_{b}SFR_{b}} &= \mathrm{SFR_{b}}/ (M_\mathrm{t}\times\mathrm{Bar/T}),
\end{aligned}
\end{equation}

where $M_\mathrm{t}$ is the total stellar mass of the galaxy that we estimate from $g$- and $i$-band magnitudes as explained in Sect. \ref{Sect:Data} and $M_\mathrm{b} = M_\mathrm{t}\times\mathrm{Bar/T}$ is the approximate stellar mass of the bar component only. To get one clean value to characterise the star formation activity in the bars, we calculated the $\mathrm{SFR_{b}}$, $\mathrm{s_{t}SFR_{b}}$, and $\mathrm{s_{b}SFR_{b}}$ within the rows $-3$ to $-1$ and $1$ to $3$ as annotated in Fig. \ref{fig:SFR_map} that we call the intermediate region of the bar. A summary of the SFRs for all galaxies can be found in Table \ref{tbl:summary_SFR}.

\begin{table*}
\caption{Summary of the SFRs from dust-corrected and AGN-masked $\mathrm{H\alpha}$ emission in the intermediate bar region.}
\label{tbl:summary_SFR}
\centering

\begin{tabular}{c c c c}
\hline\hline
Galaxy & $\mathrm{SFR_{b}}$ & $\log\mathrm{(s_{t}SFR_{b}/yr^{-1})}$  & $\log\mathrm{(s_{b}SFR_{b}/yr^{-1})}$ \\
        &       $M_\sun\,\mathrm{yr}^{-1}$      &               &\\
        (1) & (2) & (3) & (4)\\
\hline
\object{HE0021-1819}     &       $0.10 \pm 0.01$         &       $-11.11 \pm 0.11$       &        $-9.99 \pm 0.11$ \\
\object{HE0045-2145}      &       $1.87 \pm 0.19$         &       $-9.93 \pm 0.11$        &        $-9.29 \pm 0.11$ \\
\object{HE0108-4743}      &       $<1.10$         &       $<-10.60$       &       $<-9.61$ \\
\object{HE0114-0015}      &       $2.16 \pm 0.22$         &       $-10.16 \pm 0.11$       &        $-9.60 \pm 0.11$ \\
\object{HE0119-0118}      &       $3.77 \pm 0.38$         &       $-10.12 \pm 0.11$       &        $-9.21 \pm 0.11$ \\
\object{HE0253-1641}      &       $0.47 \pm 0.05$         &       $-10.93 \pm 0.11$       &        $-10.14 \pm 0.11$ \\
\object{HE0433-1028}      &       $7.72 \pm 0.77$         &       $-9.95 \pm 0.11$        &        $-9.28 \pm 0.11$ \\
\object{HE0934+0119}      &       $1.34 \pm 0.13$         &       $-9.92 \pm 0.11$        &        $-9.19 \pm 0.11$ \\
\object{HE1011-0403}      &       $0.14 \pm 0.02$         &       $-11.57 \pm 0.11$       &        $-10.80 \pm 0.11$ \\
\object{HE1017-0305}      &       $<0.11$         &       $-<11.54$       &       $<-10.74$ \\
\object{HE1029-1831}      &       $8.37 \pm 0.84$         &       $-9.44 \pm 0.11$        &        $-8.62 \pm 0.11$ \\
\object{HE1108-2813}      &       $2.53 \pm 0.25$         &       $-9.88 \pm 0.11$        &        $-9.22 \pm 0.11$ \\
\object{HE1330-1013}      &       $0.08 \pm 0.01$         &       $-11.55 \pm 0.11$       &        $-10.81 \pm 0.11$ \\
\object{HE2211-3903}      &       $0.35 \pm 0.04$         &       $-11.17 \pm 0.11$       &        $-10.40 \pm 0.11$ \\
\object{HE2222-0026}      &       $0.11 \pm 0.01$         &       $-11.07 \pm 0.11$       &        $-10.24 \pm 0.11$ \\
\object{HE2233+0124}      &       $0.16 \pm 0.02$         &       $-11.27 \pm 0.11$       &        $-10.41 \pm 0.11$ \\
\hline
\end{tabular}
\tablefoot{
(1) Galaxy name; (2) SFR in the bar region; (3) logarithm of the sSFR considering the total stellar mass of the galaxy; (4) logarithm of the sSFR considering only the stellar mass of the bar component. All measurements of star formation are integrated within the intermediate bar region (rows $-3$,$-2$, $-1$, $1$, $2$, $3$) as defined in the text and in Fig. \ref{fig:SFR_map}. The SFR in galaxy HE0108-4743 is derived from real detections, but considered as upper limit because of possible contamination from other physical processes (see Sect. \ref{Sect:nbar} for details).
}
\end{table*}

\section{Results}
\label{Sect:Results}

The purpose of this work is to investigate whether there is a connection between the presence or absence of star formation activity in the bar region and structural properties of the bar or the host galaxy. Our aim is to search for relations between the parameters that represent these properties of the bar and to analyse whether there is a clear separation into two types of bars or if we observe rather a continuous diversity.

\subsection{Star-forming versus quiescent bars}
\label{Sect:SFvsNONSF}

\begin{figure}
        \resizebox{\hsize}{!}{\includegraphics{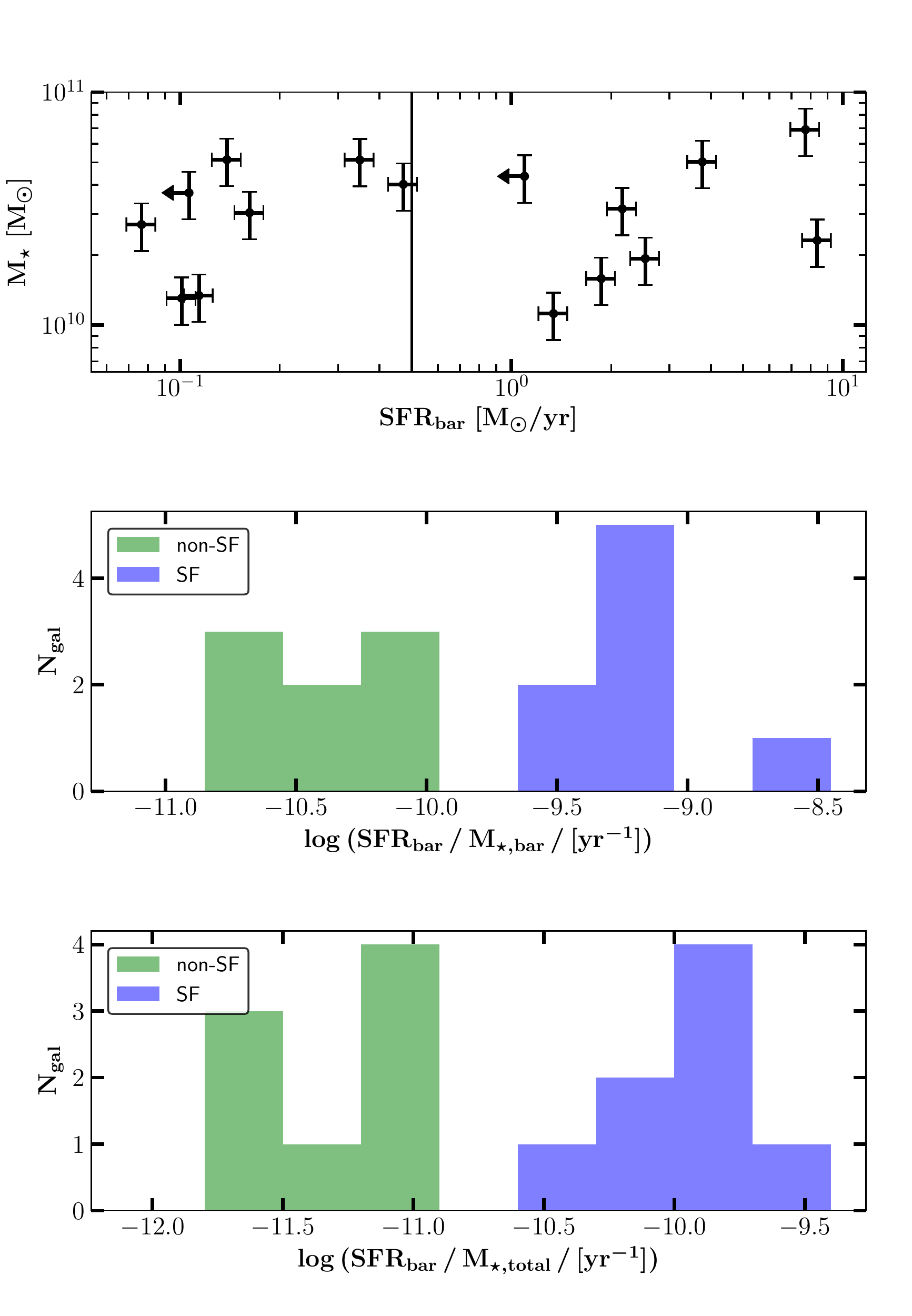}}
        \caption{Total integrated SFRs in the intermediate bar region of each galaxy that includes the rows $-3$ to $-1$ and $1$ to $3$ as defined in Fig. \ref{fig:SFR_map}. \textit{Upper panel:} $\mathrm{SFR_{b}}$ for each galaxy vs. total stellar mass. A separation between almost zero SFRs and $\mathrm{SFR_{b}} \gtrsim 1\,M_{\sun}/\mathrm{yr}$ is clearly apparent. A vertical black line at $\mathrm{SFR_{b}} = 0.5\,M_{\sun}/\mathrm{yr}$ shows our classification into SF and non-SF bars. \textit{Middle panel:} Histogram of sSFRs when accounting only for the mass of the bar. In light green and light blue we show non-SF and SF bars according to their location in the upper panel. Going from $\mathrm{SFR_{b}}$ to $\mathrm{s_{b}SFR_{b}}$ does not change the separation. \textit{Lower panel:} The same as the middle panel but dividing $\mathrm{SFR_{b}}$ by the total stellar mass of the galaxy. Again, the classification does not change.}
        \label{fig:SFR_hist}
\end{figure}

In Fig. \ref{fig:SFR_hist} we present the distribution of star formation activity in the intermediate bar region that excludes the outermost and innermost rows of the bar mask. This is to ensure we exclude contamination from spiral arms or remaining $\mathrm{H}\alpha$ emission from the AGN. The upper panel shows the SFR of each galaxy bar against the total stellar mass of the galaxy. The uncertainties of the SFRs are propagated from the errors of the emission line fluxes. A clear separation between bars with almost zero star formation and bars with clear star formation activity becomes evident in that plot. We choose $\mathrm{SFR_{b} = 0.5\ M_{\sun}/yr}$ to be the demarcation between SF and quiescent (non-SF) bars, represented by the vertical solid line. The next two panels below show histograms of the logarithm of $\mathrm{s_{b}SFR_{b}}$ and $\mathrm{s_{t}SFR_{b}}$. Both histograms independently confirm a well-defined separation into two categories of star formation activity. The absence of an intermediate population indicates that the quenching process must be quick as compared to the lifetime of SF and non-SF bars. The limited range in stellar mass may be responsible that the plot does not change much when using sSFR or SFR alone. Therefore, in our case it does not make a difference which parameter is used to divide between SF and non-SF bars, but $\mathrm{s_{b}SFR_{b}}$ might be in general the preferred discriminator.

\subsection{Bar S\'ersic index}
\label{Sect:nbar}

\begin{figure*}
        \centering
        \includegraphics[width=17cm]{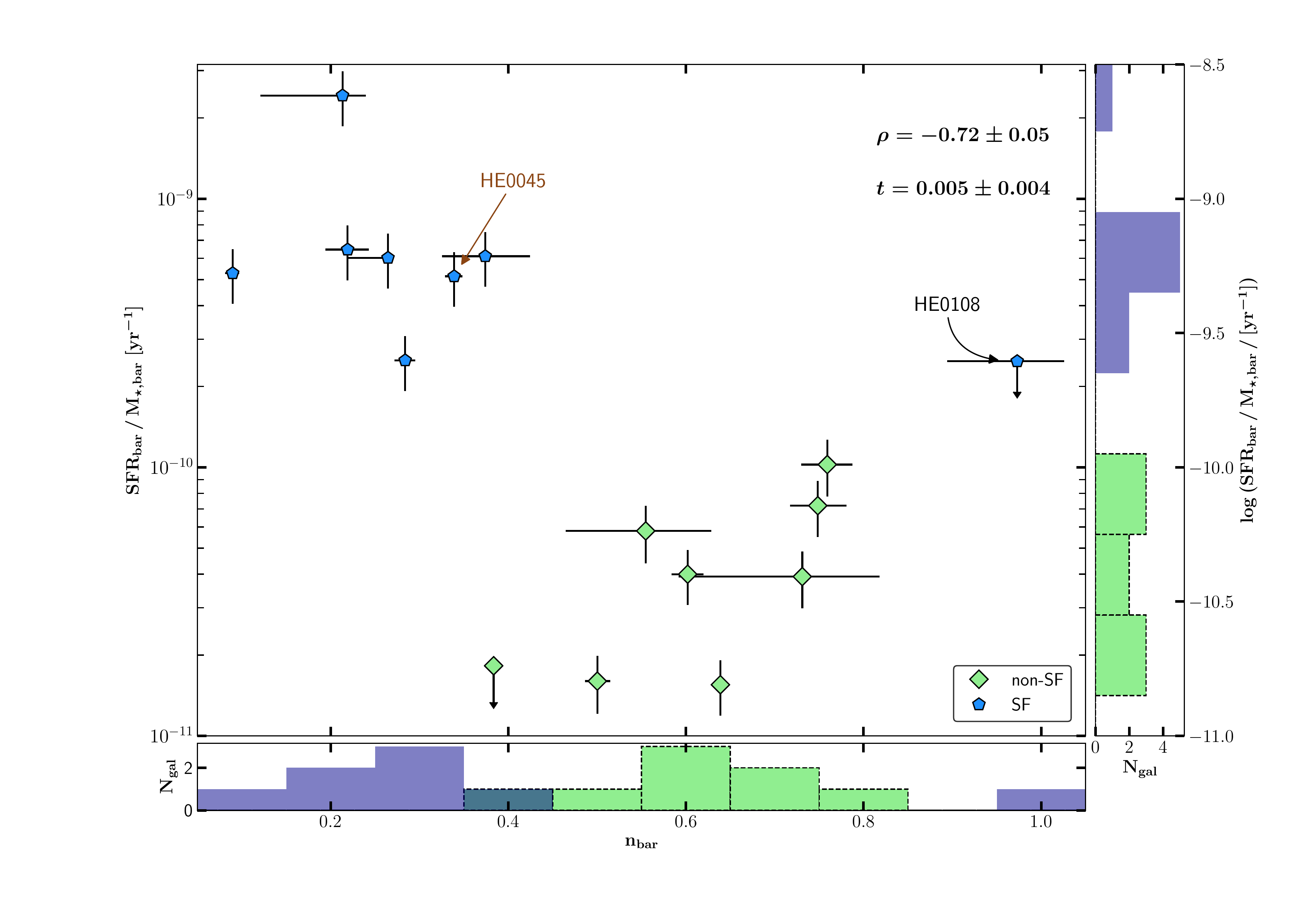}
        \caption{Comparison between the sSFR and the S\'ersic index of the bar. The $\mathrm{s_{b}SFR_{b}}$ is the SFR normalised by the stellar mass of the bar in the intermediate bar region that includes the rows $-3$ to $-1$ and $1$ to $3$ as defined in Fig. \ref{fig:SFR_map}. The index $n_\mathrm{bar}$ is the S\'ersic index of the main bar component \emph{bar1}. The objects are divided into non-SF (green) and SF (blue) according to the classification in Fig. \ref{fig:SFR_hist}. The \textit{bottom} and \textit{right panels} present histograms for both parameters. There is a clear separation in $\log\mathrm{(s_{b}SFR_{b})}$ as we have seen before. The separation is also apparent in $n_\mathrm{bar}$ with only marginal overlap. The galaxy HE0108-4743 annotated in black is discussed separately in the text. HE0045-2145 (indicated with a brown arrow) is the only galaxy not hosting an AGN in the sample. The error bars for $n_\mathrm{bar}$ show the 68\% confidence intervals from the posterior distribution of our bootstrap resampling method. More details about the uncertainties can be found in the main text and Appendix \ref{apx:nbar}. In the top right corner, we show the Spearman's rank correlation coefficient $\rho$ and the $t$-value for a null-hypothesis test.}
        \label{fig:nbar_sSFR}
\end{figure*}

\begin{figure}
        \resizebox{\hsize}{!}{\includegraphics{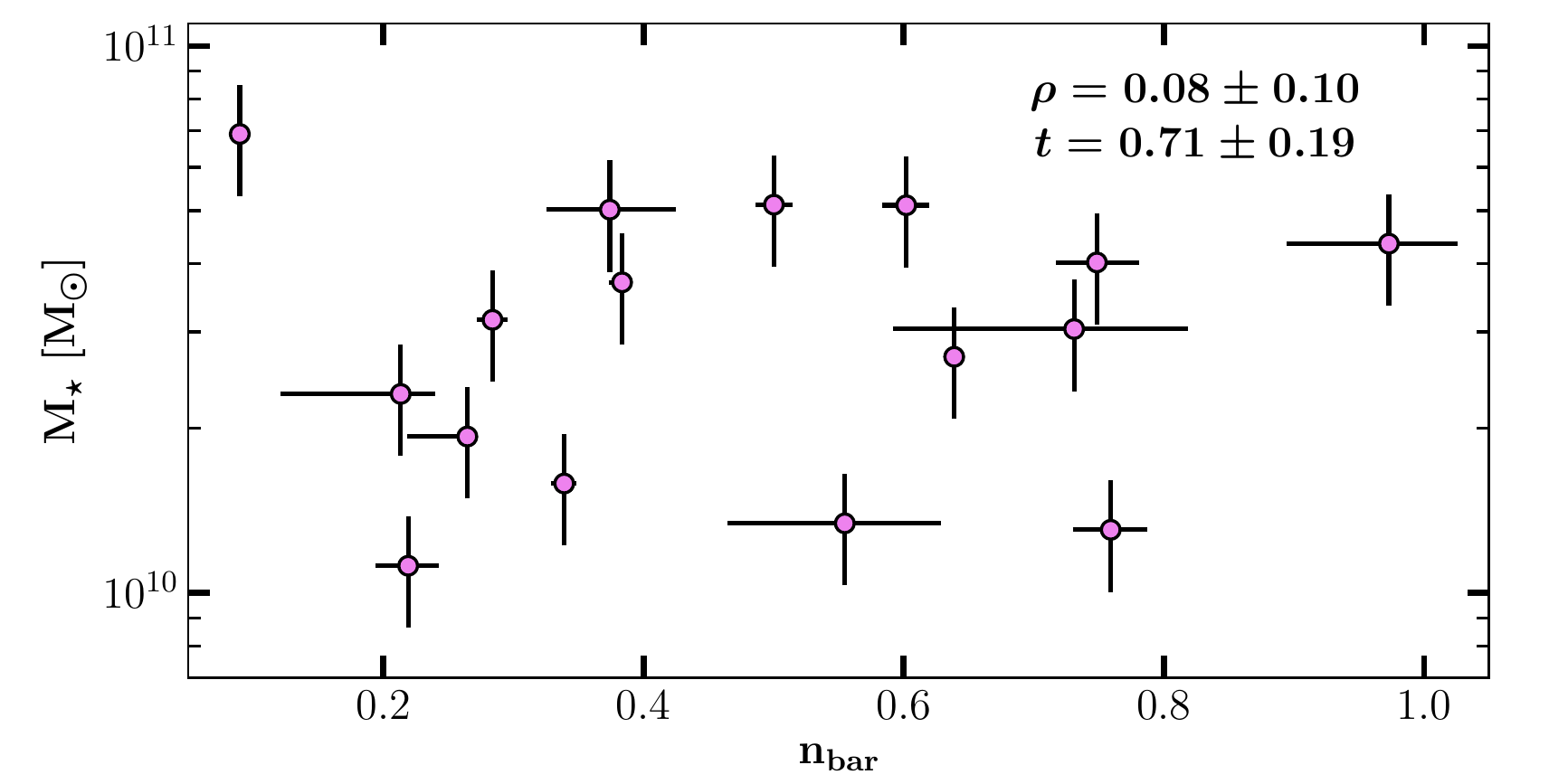}}
        \caption{Comparison between total stellar mass and bar S\'ersic index. Error bars show bootstrap errors on the x-axis and 0.1 dex uncertainties from the calibration in \citet{Taylor2011} on the y-axis. The Spearman's rank correlation coefficient $\rho$ and the $t$-value for a null-hypothesis test are given in the top right corner. The results show no indication for a correlation between the two parameters in the given mass range.}
        \label{fig:mass}
\end{figure}

Our results show no obvious correlations between star formation activity and structural parameters as for example $B/T$, $\mathrm{Bar}/T$, $n_\mathrm{bulge}$, $L_\mathrm{bar}/h$, $h$ or the type and number of parameters included in the fit, except for the bar S\'ersic index $n_\mathrm{bar}$, which is shown in Fig. \ref{fig:nbar_sSFR}. The parameter $n_\mathrm{bar}$ is the S\'ersic index of the main bar component \textit{bar1}. In this figure, we plot the specific SFR of the intermediate region of the bar against $n_\mathrm{bar}$. First of all, we observe that all bars have S\'ersic indices smaller than 1, which is typical for a bar component. A S\'ersic function with an index less than 1 produces a concave function that shows little variation in the central part and bends down towards larger radii. The smaller the S\'ersic index the sharper is the drop at the end of the bar and the flatter is the central part of the profile. The term flat is often used to describe shallow surface brightness profiles in a log-linear plot. Furthermore, the applied separation into SF and non-SF bars that we adopt from Fig. \ref{fig:SFR_hist} concurs with a trend from very low to larger S\'ersic indices, respectively. One outlier from this observed trend is discussed separately in the end of this subsection. A statistical test for a correlation between these two parameters yields a Spearman's rank correlation coefficient of $\rho = -0.72 \pm 0.05$, which clearly indicates the presence of a strong correlation. \citet{Kim2015} found that massive galaxies mainly have flat bars with $n_\mathrm{bar} < 0.4$, while less massive galaxies have close to exponential ($n_\mathrm{bar} \geq 0.8$) light profiles. Our comparison of galaxy mass and S\'ersic index presented in Fig. \ref{fig:mass} shows that all galaxies in our sample are predominantly massive ($M_\star > 10^{10}\ M_{\sun}$) and there is no correlation between $n_\mathrm{bar}$ and $M_{\star}$. The scatter in $n_\mathrm{bar}$ for the given mass range does not disagree with the presented results in \citep[][their Fig. 2]{Kim2015}. We conclude that the observed correlation of the S\'ersic index with the sSFR in our analysis cannot be explained by galaxy mass. 

In order to address the question whether the light from recently formed hot young stars in SF bars is responsible for flattening the light profile of the bar, we performed an additional multiband fit for a test case using the galaxy with the flattest and SF bar HE0433-1028. We ran a simultaneous decomposition on the collapsed $g$-, $r$-, $i$-, and $z$-band images from the MUSE cube using \textls{\sc galfitm} \citep{Haeusler2013} a modified version of \textls{\sc galfit}. The S\'ersic index did not change across the four bands within a small interval of $\Delta n_\mathrm{bar}=0.02$. Given this result, it should be safe to assume that additional light due to ongoing star formation is not a dominant cause of flat bars.

The estimation of uncertainties for parameters from 2D image decompositions is usually a difficult endeavour. Especially an increasing number of model components augments the degeneracy between the parameters and concurrently the human factor becomes more important. Most of the available codes provide a $\chi^2$ value to measure the goodness of fit. This has been shown generally to underestimate the uncertainty \citep[e.g.][]{Haeussler2007, Gadotti2009, Erwin2015a} in galaxy decompositions and, if at all, can only be used as a lower limit. To date, there is no common method that has been proven to give robust estimates of the error budget. In Appendix \ref{apx:nbar}, we discuss two different methods: one that is based on bootstrap resampling and one that follows a MCMC approach. Individual error bars for both methods can be seen and compared in Fig. \ref{fig:errors}, where we also discuss that the errors from MCMC are probably too large and the bootstrap error should be preferred. In Fig. \ref{fig:nbar_sSFR}, \ref{fig:mass}, and \ref{fig:morph} we show only the more reliable bootstrap errors. Both bootstrap and MCMC are implemented functionalities in \textls{\sc imfit}.

The galaxy HE0108-4743 stands out from the general trend found for the other galaxies in the sample. Despite a rather large S\'ersic index, it shows high star formation activity in the bar region. However, the $\mathrm{H\alpha}$ found in this galaxy is not limited to the bar region or spiral arms, but rather seems to be a continuous feature across the whole galaxy. Possibly, there has been a burst of star formation that occurred everywhere in the galaxy. This could mean that different processes are in place as compared to just the typical evolution of star formation in galaxy bars. The SFR of the bar should therefore be considered as an upper limit.

Another special case is HE2222-0026. In contrast to the other galaxies with non-SF
bars in our sample, in this galaxy we do not observe any H$\alpha$ in the outer disc and we only observe very little along the bar. Furthermore, the total molecular gas mass is rather low $M(H_2) = 0.5 \times 10^{9}\,M_\sun$ \citep{Bertram2007a}. In this special case, the absence of star formation in the bar is simply explained by a global lack of fueling gas.

\subsection{Bar surface brightness profile}
\label{Sect:loglog}

The S\'ersic function seems to yield satisfying results when it is used to fit bar components, but there are weaknesses that are important to keep in mind. The difference between S\'ersic profiles in the central flat part of functions with very low S\'ersic indices is small, as a result of which the S\'ersic index of the model is very sensitive to small variations in the data for flat bars. It is mainly governed by the strength of the cut-off in the profile, hence the end of the bar. This is problematic in the case of weak bars or the presence of inner rings or smooth transitions into spiral arms.

To test our bar fits for such cases, we chose a direct examination of the light profile as best approach. Unfortunately, the total integrated surface brightness profile confronts us with a sum of light that originates from all galaxy components. Even in the radial range where the bar dominates the light profile, it is not straightforward to recognise the light distribution of the bar only. For example, the presence of a central point source, an inner ring or strong spiral arms can alter substantially the surface brightness profile of the galaxy across a range of radii. The most appropriate solution seems to be to use our multicomponent decomposition yet having the major drawback that it already assumes certain model functions for the various components. The extent to which our assumption influences our result is determined by how much the specific bar model constrains the fit of the other galaxy components in the decomposition. 

\begin{figure*}
        \centering
        \includegraphics[width=17cm]{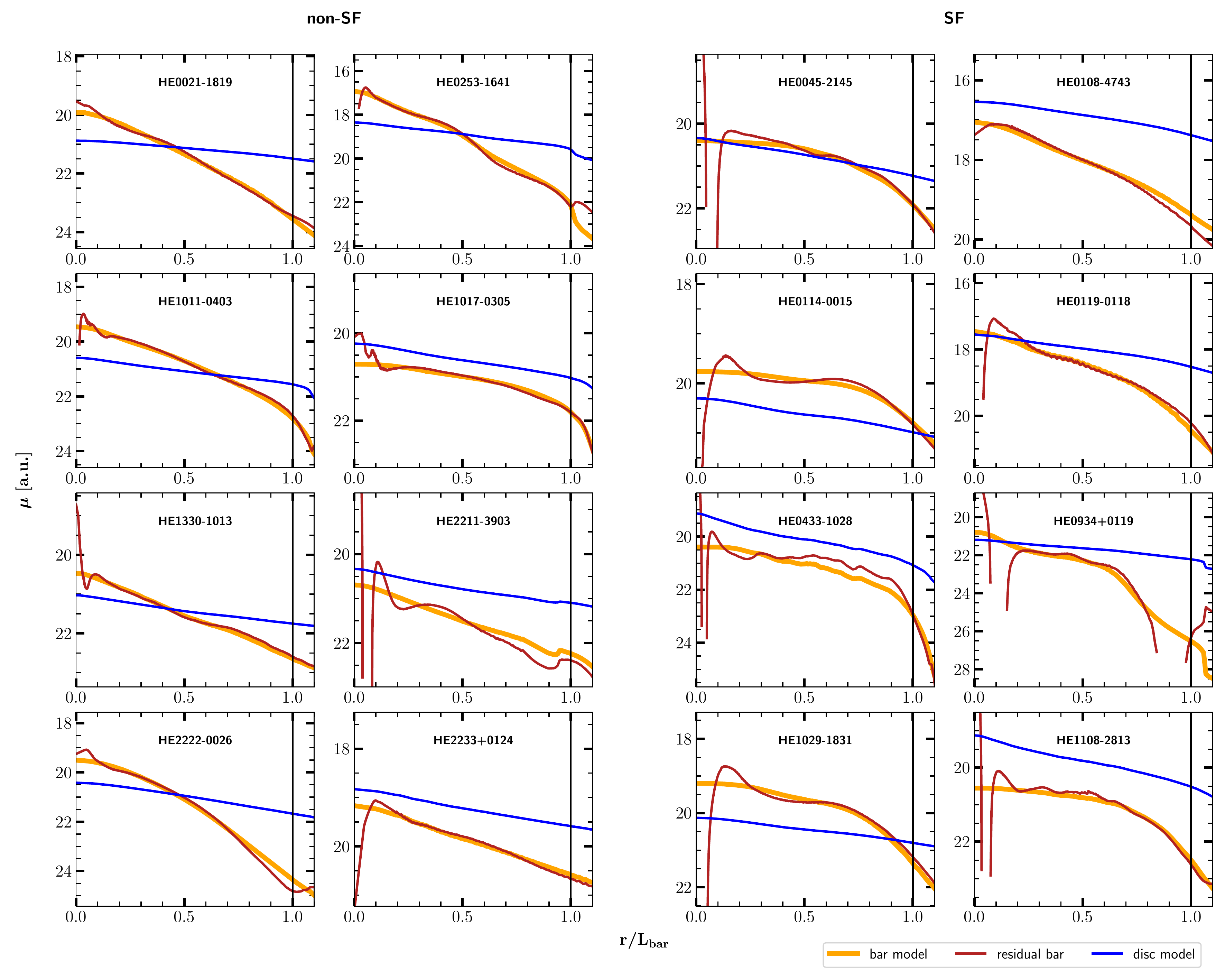}
        \caption{Residual surface brightness bar and disc profiles. On the x-axis we plot the radial distance from the centre of the galaxy along the bar major axis normalised by the bar length. The latter is shown by a vertical black line. On the y-axis we plot the surface brightness. The bar residuals are indicated in red. The orange line shows the surface brightness of the model bar component using a S\'ersic profile. The blue line shows the exponential disc profile. All profiles were extracted from the 2D image using the ellipse fitting method. The subplots are grouped according to our classification into SF and non-SF bars.}
        \label{fig:bar-profile}
\end{figure*}

We calculated the residual light profiles of the bar by subtracting all model components from the data image except the bar itself. For example, if the galaxy was best fitted with a ps+bulge+bar+disc, then we calculate $\mathrm{residual_{bar} = data_{total}-(model_{ps}+model_{bulge}+model_{disc})}$. In Fig. \ref{fig:bar-profile}, we show the bar residuals together with the disc profiles for each galaxy. For comparison, we also plot the S\'ersic model of the bar. In the case of multiple bar components in our fit, we show only the main component \textit{bar1}. All non-SF bars are on the left side and all SF accordingly on the right side.

While keeping in mind the special case of HE0108-4743 the following observations can be made.
For non-SF bars, the bar profiles are approximately exponential within the bar extent. The scale length of the bar is smaller than that of the disc. Considering the smaller scale length of the bar and a generally larger central surface brightness as compared to the disc, both profiles of the bar and the disc are in the majority of the cases crossing. 

For SF bars, the variation of the profiles with radial distance is generally higher than among the non-SF bars. The inner parts tend to be flatter with a scale length of similar size to that of the disc. Further out, the profiles show a sharp drop that usually starts before the measured end of the bar. 

In summary, the surface brightness profiles of SF bars are similar to those of their corresponding discs in the inner 50--80\% of their lengths, where they reach a drop-off point. In contrast, non-SF bars have much smaller scale lengths with no clear down bending at the end or near the end of the bar. Compared to the bar S\'ersic index, we note that both methods are able to make the same distinction between both classes of objects. However, the results of $n_\mathrm{bar}$ give the impression that non-SF bars are more similar to their discs by having close to exponential S\'ersic indices; yet the direct observations of the profiles yield contrary conclusions since their scale lengths are substantially shorter than the scale lengths of their host discs. All in all, $n_\mathrm{bar}$ can be used as a measure of the flatness of the bar, but a bar with a close-to-exponential index should not be considered to be more similar to the disc than a bar with a smaller index.

\subsection{Morphology of host galaxy}
\label{Sect:morph}

Some previous studies have reported that the aforementioned bar features correlate with the morphological type of the host galaxy. Early-type disc galaxies have been found to have flat \citep[e.g.][]{Elmegreen1985, Elmegreen1996, Ohta1996g} and non-SF bars \citep[e.g.][]{Phillips1996}. In contrast, late-type discs have supposedly exponential and SF bars.

\begin{figure}
        \resizebox{\hsize}{!}{\includegraphics{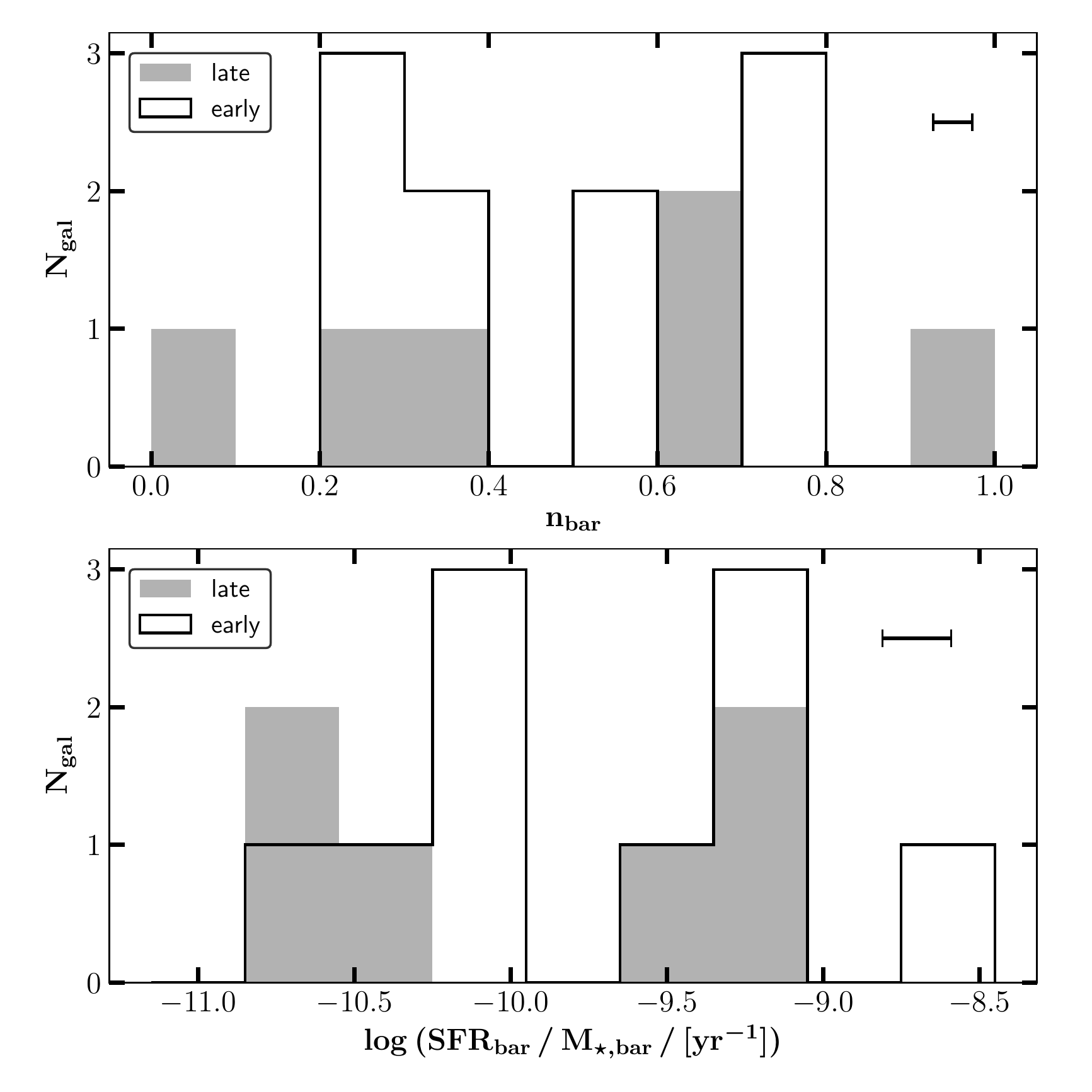}}
        \caption{Morphology of the host galaxies. Comparison between early-type (Hubble type $T\leq3$, 10 objects) and late-type ($T>3$, 6 objects) spiral galaxies. \textit{Upper panel:} Distribution of the S\'ersic index of the primary bar component $n_\mathrm{bar}$. \textit{Lower panel}: Distribution of the sSFR in the bar region $\mathrm{s_{b}SFR_{b}}$. The median errors of $n_\mathrm{bar}$ and $\log\mathrm{(s_{b}SFR_{b})}$ are shown in the upper right corner. A typical error of Hubble type classifications for spiral galaxies is $\sigma=1.01$ T-types \citep[based on >700 classification of 5 observers,][]{Walcher2014}. A k-sample Anderson-Darling test yields that the hypothesis that the early- and late-type galaxy samples are drawn from the same parent distribution cannot be rejected for either the S\'ersic index nor $\mathrm{s_{b}SFR_{b}}$ with significance levels of 92\% and 55\%, respectively.}     
        \label{fig:morph}
\end{figure}

We performed a classification into Hubble types (see Sect. \ref{Sect:Data} and Table \ref{tbl:summary_decomp}) and compared them with the parameters $n_\mathrm{bar}$ and $\mathrm{s_{b}SFR_{b}}$ in Fig. \ref{fig:morph}. The distributions show no significant differences between early-type (Hubble type $T\leq3$, 10 objects) and late-type ($T>3$, 6 objects) spiral galaxies for either of the parameters. A k-sample Anderson-Darling test shows that they are consistent with being drawn from the same parent population.

Our sample comprises four galaxies with an inner ring (HE0021-1819, HE1017-0305, HE1330-1013, and HE2211-3903). An inner ring is usually defined as the ring-like structure that lies just outside the bar. Interestingly, all four galaxies are non-SF within the bar, while they show some star formation along the ring. However, not all non-SF bars co-exist with inner rings. Inner rings may be connected to the star formation activity in the bar (or absence thereof), but their presence is not a necessary condition for non-SF bars.

\subsection{Spatial distribution of star formation}
\label{Sect:results_SFdist}

\begin{figure}
        \resizebox{\hsize}{!}{\includegraphics{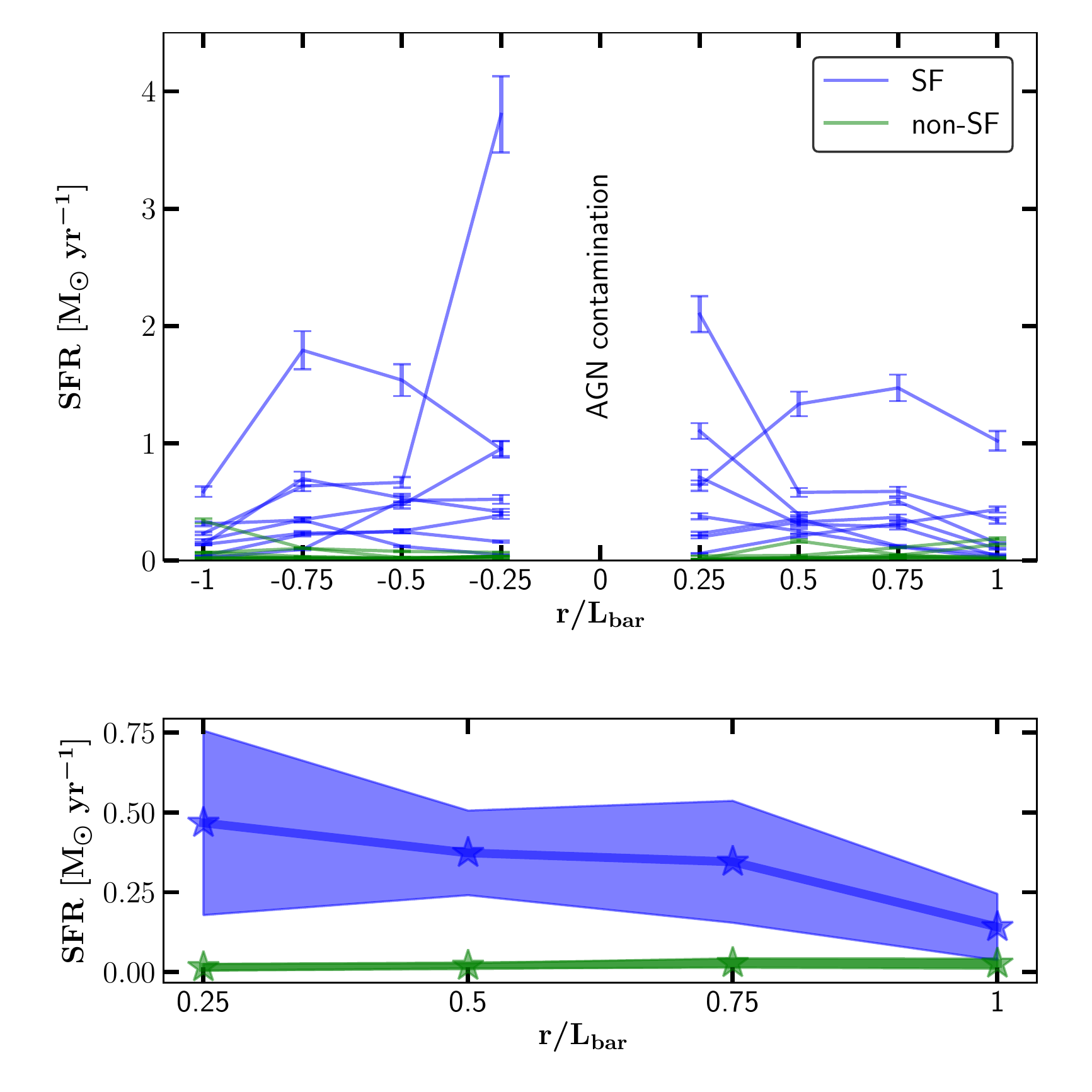}}
        \caption{\textit{Upper panel:} Distribution of the SFRs across the bar parallel to the bar major axis. Each data point shows the integrated SFR (y-axis) within the corresponding row (x-axis) in the bar mask as indicated in Fig. \ref{fig:SFR_map}. Each line corresponds to one object. The SF bars are shown in blue and non-SF bars are indicated in green. The central row is not shown since the SFR is very uncertain, because of over  or under subtraction of AGN contamination. \textit{Lower panel:} Median of both sides of the bar and all objects within each category. The shaded regions show the median average deviation (MAD). On average, the SFR of SF bars is decreasing with distance.}
        \label{fig:SFR_radial}
\end{figure}

In addition to distinguishing between star formation that is happening inside or outside the bar, we can also pinpoint the direct location (in 2D projection) of SF sites within the bar. In Fig. \ref{fig:SFR_map} we showed the bar mask used for this measurement. Instead of using the total sum of star formation in the bar, it is also possible to plot a radial profile of SFR for each galaxy. Fig. \ref{fig:SFR_radial} shows the SFR along the bar major axis (row number $-4$ to $4$ in Fig. \ref{fig:SFR_map}). Each bin shows the sum of the three columns $-1$ to $1$, it is therefore not a cut across the major axis, but contains the whole SFR within the width of the bar.

In addition to the applied method to disentangle H$\alpha$ flux from AGN and star formation, we also excluded the central bin from this plot. We do not observe any significant difference between both sides of the bar. On average, the figure shows a trend of decreasing SFR with radial distance for SF bars, although the scatter is large and some individual objects show the opposite trend. The cause of that could be an effect of gas density or other local variables. This trend agrees with findings of radial decreasing numbers of core-collapse supernovae \citep[e.g.][]{Hakobyan2009,Herrero-Illana2012}, which have been found to follow H$\alpha$ distributions in SF galaxies \citep{Anderson2009}.

\begin{figure}
        \resizebox{\hsize}{!}{\includegraphics{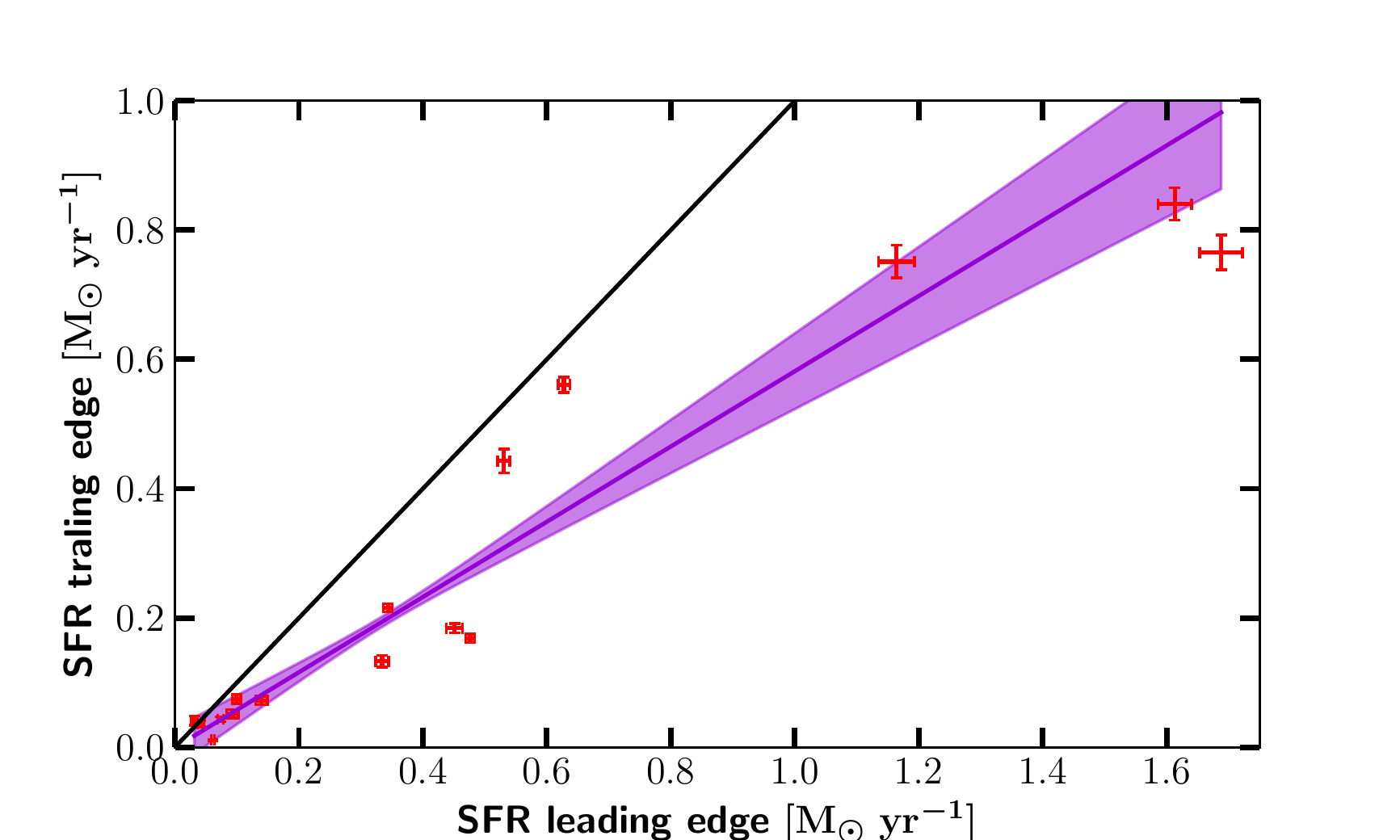}}
        \caption{Comparison between the SFR on the leading and on the trailing edge of the bar. The black solid line shows a hypothetical one-to-one correlation. The data suggest a linear correlation that was tested with the Pearson correlation coefficient that yields $\rho = 0.96 \pm 0.01$. The violet solid line shows a linear regression of the form $y = m\cdot x + c$ with the parameter estimates $m = 0.58 \pm 0.09$ and $c = 0.00 \pm 0.01$. On average, the SFR is stronger on the leading edge of the bar by a factor of 1.76.}
        \label{fig:SFR_minor}
\end{figure}

A look at the distribution across the three columns of the mask (going perpendicular to the bar major axis) helps us to understand whether the star formation is preferentially happening either on the leading or the trailing edges of the bar. We define the direction of rotation of the bar under the assumption that spiral arms are trailing in our galaxies. In Fig. \ref{fig:SFR_minor} we plot the SFR of the leading edge against the SFR of the trailing edge. We recognise that for all but one bar the star formation is stronger on the leading edge. There is an even strong indication for a correlation between the SFR on both edges. A Pearson test for linear correlation yields a correlation coefficient of $\rho = 0.96 \pm 0.01$. The result of a linear regression is given by $\mathrm{SFR}_\mathrm{trailing} = (0.58 \pm 0.09)\times \mathrm{SFR}_\mathrm{leading} + (0.00 \pm 0.01)$. 
The prevalence of star formation on the leading edge of the bar is in good agreement with previous observations \citep[e.g.][]{Sheth2002a} and simulations \citep{Athanassoula1992, Renaud2015b}. Observationally, molecular gas clouds are predominantly found on the leading edges. This has been confirmed in simulations by converging flows and large-scale shocks that yield higher gas densities. Additionally, \citet{Renaud2015b} found that tangential velocity gradients are less strong on the leading edge that makes it easier for the gas clouds to collapse and form stars. The spiral-like pattern that can be seen in the SFR maps in Fig. \ref{fig:SFR_map} and Figs. \ref{fig:sfr_map1} to \ref{fig:sfr_map3} is typical. This pattern follows the leading edges of the bar (frequently traced by dust lanes), which is built inside corotation from the \emph{x1} orbits parallel to the bar. If there are two inner Lindblad resonances (ILR) there are \emph{x2} orbits perpendicular to the bar, and a possible nuclear bar inside a nuclear ring, to trigger further gas infall to the centre of the galaxy. (\citealp{Athanassoula1992}; \citealp[see also][]{KimWT2012,Sormani2015,Fragkoudi2016}).

\section{Discussion}
\label{Sect:discuss}

\subsection{Comparison with previous works}

\subsubsection{Flatness of the bar}
\label{Sect:elmegreen}

In \citet{Elmegreen1985}, the authors classified 15 barred galaxies into bars with flat and exponential-like surface brightness profiles based on $I$-band surface photometry on photographic plates. A bar is described as flat if the major axis profile is flatter than the outer profile along the spiral arm and it is exponential if it is similar or even steeper than the spiral profile. They further subdivide the flat bars into a group of bars with flat interbar intensity profiles and a group with much faster interbar intensity decrease. When compared to Hubble types, flat bars are preferably found in early-type spiral galaxies, while late-type spirals tend to have exponential bars. These findings were confirmed in \citet{Elmegreen1996} with a (partially overlapping) sample of 19 barred galaxies in the $B$, $I$, $J,$ and $K$ band on CCD detectors. In the latter work, they do not use the spiral arm profiles, but a straight continuation of the major axis profiles into the disc. They also plot minor axis profiles instead of interbar averages. The main result is that flat bars are located in SBb-SBc galaxies and exponential bars in  SBc-SBm types.

In this work, we do not find a correlation between Hubble type and flatness of the bar. Our estimate of the flatness is based on the S\'ersic index of the main bar component from a 2D photometric decomposition. A comparison of this estimate with residual bar light profiles in Sect. \ref{Sect:loglog} confirmed our classification. In addition, for the purpose of a cleaner comparison to \citet{Elmegreen1985}, we constructed major and minor axis light profiles for our sample shown in Fig. \ref{fig:majorminor}. By purely examining the major axis profiles, we find only four bars that are exponential-like, i.e. HE0021-1819, HE0108-4743, HE2222-0026, and HE2233+0124. The S\'ersic indices of these bars are between $0.55 \leq n_\mathrm{bar} \leq 0.97$, thus, they are among the larger values. Their Hubble types range from SBa to SBc. All other bars have flatter profiles. This analysis confirms our result that there is no preferred Hubble type for a flat or for an exponential bar. Both types of bars occur from SBab to SBc. However, we do not cover types later than SBcd.

Given the limited sample sizes and the subjectivity of Hubble type classification, we refrain from claiming global statements, but with our detailed analysis of surface brightness profiles of galaxy bars using high quality data, we caution that a relation between the flatness of the bar and Hubble type is not as simple as hitherto thought.

\subsubsection{Star formation within the bar}

A correlation between Hubble type and star formation along the bar was reported in \citet{Phillips1993k}, \citet{Phillips1996} and \citet{Garcia-Barreto1996}. In these works the authors use narrowband $\mathrm{H}\alpha$ images to analyse the distribution of SF sites in the galaxies. The sample in \citet{Phillips1993k} comprised 15 barred spiral galaxies of SBb to SBc Hubble types. He found that SBb galaxies have moderate to no star formation along the bar, while SBc galaxies have luminous \ion{H}{II} regions within the bar. In \citet{Phillips1996}, in addition to the previously mentioned sample the author included $\mathrm{H}\alpha$ observations from the literature (without precise specification), concluding that galaxies of SBb and earlier types show no star formation in the bar, whereas bars in SBbc and later galaxies are actively SF. Similarly, \citet{Garcia-Barreto1996} found that 18 out of 52 barred spiral galaxies in their sample show star formation within the bar. Five of these 18 galaxies are SBb or earlier and 13 galaxies are SBbc or later.

In this work, we conducted a detailed measurement of SFRs based on dust-corrected and AGN-star formation deblended H$\alpha$ measurements in spatially resolved, well-defined regions within the bar. Our Fig. \ref{fig:morph} shows no correlation of star formation activity in the bar with Hubble type. In fact, galaxies of the same type can host both SF and non-SF bars. A nice example shows the comparison of HE1108-2813 with HE1017-0305. Both galaxies are classified as SBc, yet the former is actively SF ($\mathrm{SFR_{b}} = 2.53\,M_\sun\,\mathrm{yr^{-1}}$) within the bar and the latter is not ($\mathrm{SFR_{b}} < 0.11\,M_\sun\,\mathrm{yr^{-1}}$). This can clearly be seen already in the H$\alpha$ contours in the images in Fig. \ref{fig:all_images}. In summary, we show that bars of all Hubble types between SBa and SBcd can be SF and non-SF.

\subsection{AGN feeding}
\label{Sect:AGNfeeding}

Simulations have succesfully shown that the gravitational potential of a large-scale bar induces gas inflow towards the centre of the galaxy through torques and angular momentum transfer. Thereby they are able to provide the fuel for nuclear activity \citep{Shlosman1989, Shlosman1990, Piner1995, Fragkoudi2016}. Such inflows have been seen observationally in \citet{Holmes2015}, who finds non-circular flows in H$\alpha$ velocity fields in 12 out of 29 galaxies with 11 of 12 stemming from bars. 

Observational evidence trying to link the presence of a bar to AGN activity is still not conclusive. To test the scenario of bar-driven AGN feeding, studies have compared the incidence of bars in active and non-active galaxies as well as the incidence of AGN in barred and unbarred galaxies. While some of these studies report results that support the hypothesis that bars fuel AGN \citep{Knapen2000, Coelho2011, Oh2012, Alonso2014, Galloway2015}, others find no significant correlations \citep{Mulchaey1997, Ho1997, Cheung2015}. \citet{Oh2012} point out the importance of breaking degenerate correlations between bar effects and galaxy properties.

Additionally, the influence of bars on AGN has also been studied by measuring the strength of the nuclear activity as compared to the presence and strength of the bar. \citet{Cisternas2013} studied the activity of low-luminosity AGN in 41 barred host galaxies with Chandra X-ray observations and near-infrared Spitzer data and found no correlation between nuclear activity and bar strength, irrespective of galaxy luminosity, stellar mass, Hubble type, and bulge size. \citet{Goulding2017} approached the problem that AGN activity changes on much smaller timescales compared to the lifetime of bars by stacking Chandra X-ray sources as proxy for a time-averaged accretion. These authors also concluded that bars have little or no effect on the nuclear activity. Another possible explanation for a lack of correlation in some studies apart from different timescales is provided by \citet{Fragkoudi2016} who found in their simulations that boxy/peanut bulges reduce the gas inflow rate by more than an order of magnitude. By contrast, in an extensive study of $\sim 5000$ AGN \citet{Alonso2018} have found that AGN in barred galaxies show an excess of nuclear activity (as measured from $L_{[\ion{O}{iii}]}$) and accretion rate as compared to AGN in unbarred galaxies.

In this context, it is interesting to investigate whether the capability of a bar to  transport gas to the centre and fuel AGN depends on bar properties other than the bar strength. The non-SF and SF bar types in our sample might reflect the dynamical state and age of the bar or its gas content. In the following we test if the star formation activity in the bars of our sample correlates with nuclear activity.

\begin{figure}
        \resizebox{\hsize}{!}{\includegraphics{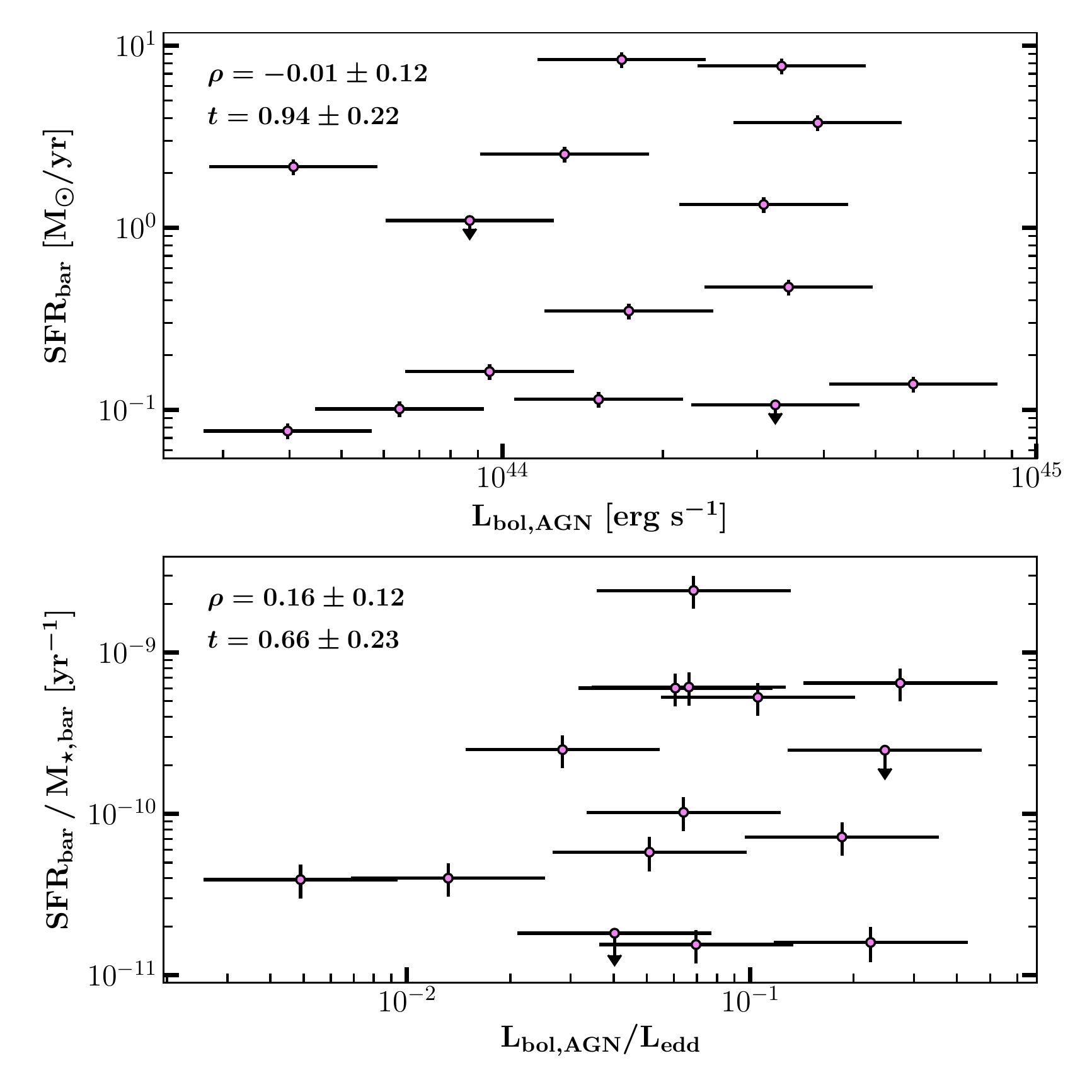}}
        \caption{Testing AGN feeding: comparison between star formation activity in the bar and nuclear activity in the host galaxy. \emph{Upper panel:} SFR in the bar region against bolometric luminosity of the AGN. \emph{Lower panel:} sSFR in the bar against Eddington ratio. The Spearman's rank correlation coefficient $\rho$ and its $t$-value for a null-hypothesis are given in the top left corner of each panel. HE0045-2145 is not included because it does not host an AGN.}
        \label{fig:AGN_lum}
\end{figure}

In the upper panel of Fig. \ref{fig:AGN_lum} we plot $\mathrm{SFR_{b}}$ against the AGN bolometric luminosity ($L_\mathrm{bol}$) which we derive from the host-subtracted (see AGN-host deblending procedure, explained in Sect. \ref{Sect:SFR}) monochromatic luminosity $\lambda L_\lambda$ (5100 $\AA$) at rest-frame wavelength of $5100\ \AA$ following \citet{Wandel1999} and \citet{Kaspi2000}. The lower panel shows $\mathrm{s_{b}SFR_{b}}$ against the Eddington ratio ($L_\mathrm{bol}/L_\mathrm{edd}$). There is no indication for a correlation between these parameters in either of the two plots. However, the restricted range in AGN luminosities ($\sim$ 1 dex) and Eddington ratios ($\sim$ 1.5 dex) limits an optimal comparative study. Although bars may enhance the nuclear activity, the effect of SF and non-SF bars does not show clear differences. Nevertheless, a causal connection between bars and AGN activity is difficult to observe owing to the different timescales of long-living stellar bars, variable AGN activity, and tracers of ongoing star formation such as H$\alpha$.

\subsection{Implications on the evolution of stellar bars}

Stellar bars form spontaneously from instabilities in the galactic disc \citep[e.g.][]{Athanassoula2013a}. It seems reasonable to assume that as they form from the disc material, they start with similar properties as the underlying disc in terms of radial surface brightness profile (exponential) and ongoing star formation activity (SF) in the early stages of their formation and evolution. As they evolve, bars grow stronger and longer and eventually must stop forming stars because of depletion of gas, as long as there is no external replenishment. Additionally, other processes such as shear may be in place that restrict star formation locally and accelerate the quenching process in the bar. This scenario can be seen for example in the simulation of \citet[][]{Khoperskov2018}.

In the present paper, we show that the flattest bars ($n_\mathrm{bar} \lesssim 0.4$) -- those with very shallow profiles in the log-linear radial surface brightness diagram -- are actively SF and bars that are less flat ($0.4 \lesssim n_\mathrm{bar} \lesssim 0.8$) are non-SF. If bars evolve from one type into the other, this suggests either that young bars have shallower profiles that grow steeper over time or that non-SF bars can start forming stars at later stages of their evolution. The non-correlation of SFR in the bar with $B/T$ and Hubble $T$-type can be explained by a variability of gas in the bar during the evolution.

If we look at the parameters of our sample, we see that this study is missing low-mass galaxies $\left(M_\star < 10^{10}\ M_\sun\right)$, exponential bars $\left(n_\mathrm{bar} > 0.8\right),$ and  very late Hubble types (SBcd-SBm). Exponential bars have been connected to low-mass galaxies \citep{Kim2015} and late-type spirals \citep{Elmegreen1985}. Late-type galaxies of low stellar masses with exponential bars, which have been associated with star formation in the bar, could be an important population of bars to complement the evolutionary picture. Nevertheless, our results indicate that the SF bars in our sample are not part of that population. In other words, earlier-type spirals in massive galaxies can host flat and SF bars.

Another important part of the puzzle is the presence of inner rings. Ring-like structures in disc galaxies are often associated with dynamical orbital resonances; but see also \citet{Romero-Gomez2006,Romero-Gomez2007} in which the manifold theory is employed to explain such structures. The most important resonances in a barred galaxy are the outer Lindblad resonance (OLR), the corotation resonance (CR) and the inner Lindblad resonance (ILR). Usually the OLR is located at about twice the bar length, the CR just outside the bar region and the ILR well within the bar region \citep[e.g.][]{Comeron2014}. The non-axisymmetric potential of a barred galaxy induces gravitational torques that drive gas outside the CR outwards where it accumulates at the OLR. Gas inside the CR is driven inward towards the ILR and/or the inner 4/1 resonance, which is also called ultra-harmonic resonance (UHR) just inside corotation \citep{Schwarz1984a,Buta1996}. This resonant accumulation of gas can lead to the formation of outer rings at the OLR, inner rings between the UHR, and the CR and nuclear rings at the ILR \citep[see also][]{Buta1986,Rautiainen2000}

All four galaxies with inner rings in our sample host non-SF bars. However, not all non-SF bars host inner rings (see Sect. \ref{Sect:morph}). This indicates that the conditions for a galaxy to quench star formation in the bar might be necessary, but they are not sufficient conditions to develop an inner ring. It could also be a timescale effect. During the evolution of the bar it depletes itself from gas, hence the star formation decreases, while at the same time gas accumulates near the UHR fueling an inner ring. After the formation of the ring it becomes more difficult to replenish the bar region with gas, especially if the galaxy is relatively isolated. This scenario is in agreement with the reciprocal relationship of \ion{H}{II} regions between the bar and an inner ring found by \citet{Ryder1993a}. A similar relationship should also be observable between star formation in bars and nuclear rings and would be an interesting subject for a future work.

\section{Conclusions}
\label{Sect:Conclusions}

We have used MUSE/VLT IFS data from the CARS survey to accurately photometrically decompose a sample of 16 local barred spiral galaxies including up to 7 different components. We additionally derived spatially resolved SFRs from dust-corrected $\mathrm{H\alpha}$ line fluxes and analysed the total amount and the spatial distribution of star formation within the bar component. From a detailed comparison of the obtained parameters, we summarise the following conclusions:

\begin{enumerate}
\item There are two classes of galaxy bars: those that show significant star formation (SF bars) and those that have very little to no star formation activity (non-SF bars). A third category of bars with fading star formation in the centre as proposed by \citet{Verley2007c} could not be probed because of AGN dominated photoionisation in the central region. The clear separation in $\mathrm{s_{b}SFR_{b}}$ between SF and non-SF bars indicates that the quenching process must be fast as compared to the lifetime of SF and non-SF bars.

\item The $\mathrm{SFR_{b}}$ and $\mathrm{s_{b}SFR_{b}}$ correlate with the S\'ersic index $n_\mathrm{bar}$ of the main bar component; in fact, we observe a separation between SF and low-index ($n_\mathrm{bar} \lesssim 0.4$) bars and non-SF and high-index ($0.4 \lesssim n_\mathrm{bar} \lesssim 0.8$) bars.

\item We find that SF bars are flatter and have profiles that have a similar slope to that of the underlying disc up to a radius where the brightness suddenly drops, whereas non-SF bars have closer to exponential profiles with a smaller scale length than the disc and no clear downturn within the bar length. The flatness of a bar is a term that has been used in the literature to describe the surface brightness profile of some stellar bars. It is, however, misleading since a flat bar can still be exponential, but with a larger scale length, up to the radius where it turns down. Flat bars might actually be more similar to their discs than exponential bars.

\item There is no significant difference in the distribution of sSFR of the bar ($\mathrm{s_{b}SFR_{b}}$) or bar S\'ersic index ($n_\mathrm{bar}$) between early-type and late-type disc galaxies. Both earlier and later types can have star formation or not and can be flat or exponential. This is in contrast to previous reports \citep{Elmegreen1985,Elmegreen1996,Ohta1996g,Phillips1993k,Phillips1996}. 
Compared with this literature the range of Hubble types of our sample is similar and we only miss the very late types SBd to SBm, which are however the types previously reported to host exponential bars. Most of the bars in our sample are indeed rather flat. Furthermore, neither the samples in the aforementioned literature nor our sample are of statistically significant size to make global statements, but -- given the depth of our analysis -- this is a cautionary note: a plain correlation between Hubble type and flatness or star formation activity might be too simple. Early-type spirals in massive galaxies can host flat and SF bars.

\item The radial distribution of SFR of SF bars is decreasing with increasing distance from the centre.

\item Star formation activity is about 1.75 times stronger on the leading edge than on the trailing edge of the bar, in good agreement with previous works \citep[e.g.][]{Athanassoula1992,Sheth2002a,Renaud2015b}.

\item The presence of non-SF bars might be related to the presence of inner rings.

\item The $\mathrm{SFR_{b}}$ is not correlated with the bolometric luminosity of the AGN, nor is $\mathrm{s_{b}SFR_{b}}$ correlated with the Eddington ratio. Hence, there is no evidence that the star formation activity in the bar affects AGNs feeding. However, given the potential unknowns of the effects of selecting luminous type-1 AGN and the therefore restricted range of AGN luminosities and Eddington ratios, the conclusions from this work may only apply to type-1 AGN hosts. Further work is required to confirm whether they can be extended to the full population of barred galaxies.

\end{enumerate}

Our analysis is by construction based on a sample of AGN host galaxies, which raises the question whether the presence of the AGN affects the SFRs in the bars measured in this work. There is no obvious answer to that. Not only different and uncertain timescales when AGN feedback becomes visible, but also the radial range it affects is still under debate \citep[e.g.][]{Gaspari2017b, Harrison2017}. The impact on the results presented in this work is frankly unknown. One of the galaxies, HE0045-2145, which was first misclassified to host an AGN, does not show exceptional results in any way in our analysis. We suggest a comparable study with an AGN-free control sample in the future, but until then there is no reason to believe that barred galaxies without AGN would yield different results. Further papers from the CARS collaboration will address the effect of feedback in the future.

\begin{acknowledgements}

We thank the anonymous referee for the constructive feedback that helped to improve this paper. We thank Darshan Kakkad, Peter Erwin, and Bruce Elmegreen for useful comments on method and content. TAD acknowledges support from a Science and Technology Facilities Council Ernest Rutherford Fellowship. M.G. is supported by the \textit{Lyman Spitzer Jr.} Fellowship (Princeton University) and by NASA Chandra GO7-18121X and GO8-19104X. MK acknowledges support by DLR 50OR1802.

\end{acknowledgements}



\bibliographystyle{aa}
\bibliography{NewDatabase}



\begin{appendix}
\onecolumn

\section{Uncertainties of the 2D photometric decomposition}
\label{apx:uncertainties}

\subsection{Uncertainty of $\lowercase{n_\mathrm{bar}}$}
\label{apx:nbar}

\begin{figure*}[b]
        \centering
        \includegraphics[width=17cm]{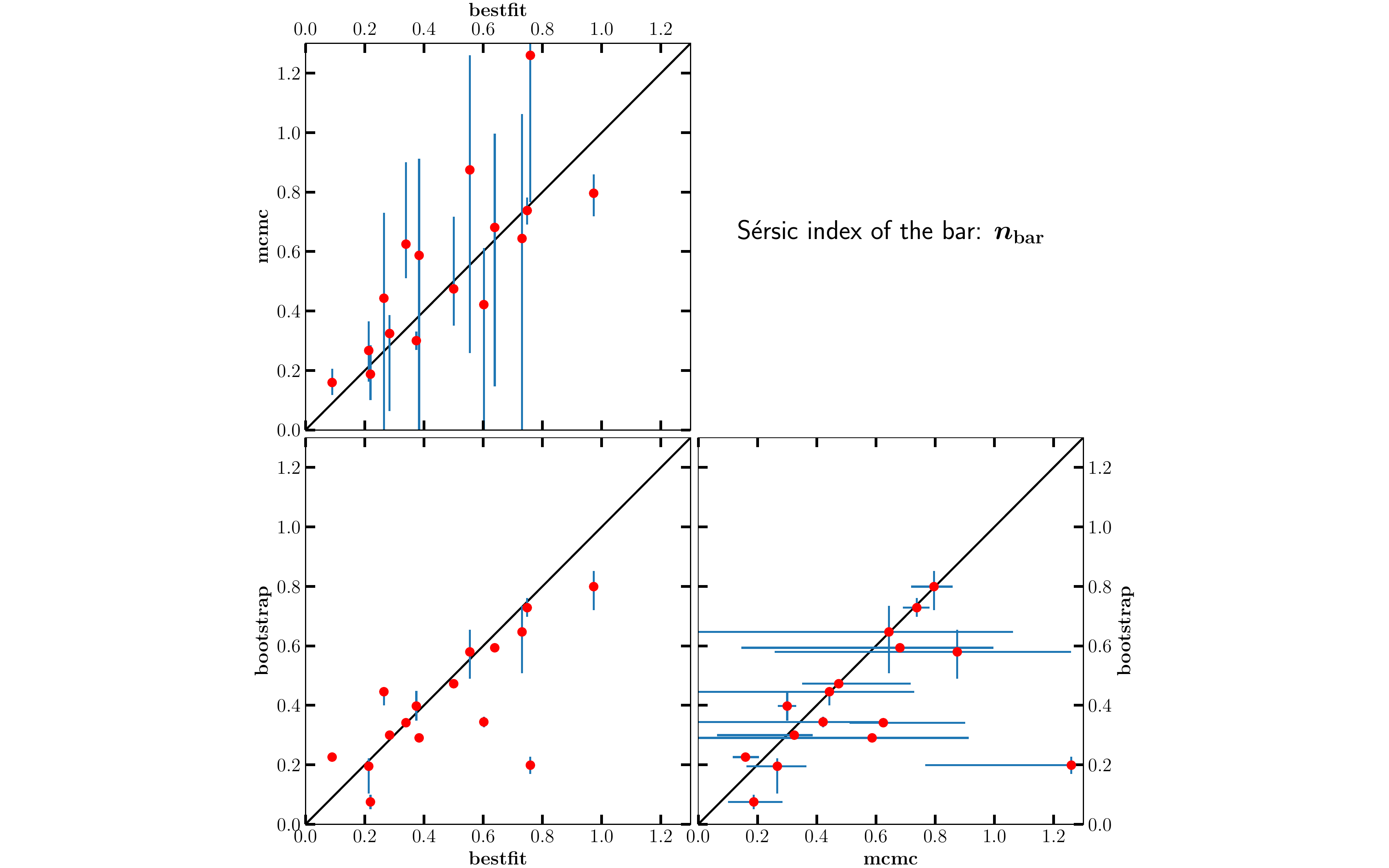}
        \caption{Comparison between different methods to fit multicomponent galaxy models to the collapsed 2D MUSE $i$-band images. This plot only shows results for the parameter $n_\mathrm{bar}$. The label \textit{best fit} is the best fitting result using the Nelder-Mead simplex minimisation technique and a $\chi^2$ fit statistic for minimisation. The label \textit{bootstrap} gives the result of resampling the pixel values in the data image with 1,000 iterations after the best fit. For time-saving reasons it uses the Levenberg-Marquardt minimisation algorithm. The red data points given in the plot indicate the median value of the 1,000 fits; the error bars show the 0.16 and 0.84 quantiles.\ The label \textit{mcmc} is the result of applying a MCMC analysis to the data image instead of searching for the best fit model. The data points in the plot indicate again the median value of the posterior distribution; the error bars show the 0.16 and 0.84 quantiles.}
        \label{fig:errors}
\end{figure*}

In this subsection, we discuss two different approaches to estimate the uncertainties of 2D image decompositions; both are implemented in the code \textls{\sc imfit}. The first method uses a bootstrap technique after finding the best fit result. Per iteration, it resamples and replaces a fraction of the data points and repeats the fit using the fast Levenberg-Marquardt minimisation algorithm given the best fit parameter values as initial guesses. We perform the bootstrapping with 1,000 iterations and take the inner 68\% range of the posterior distribution as error. The other approach is based on a MCMC analysis. It starts with multiple walkers that explore the parameter space that is given by the user as input. It is completely independent from the best fit result. We choose to equal the number of walkers to the number of free parameters (which range between 11 and 34 depending on the number of model components in the fit) as suggested in the documentation of the code; we used a 5,000 step burn-in phase and an upper limit of 100,000 steps to terminate the script if no convergence is reached. Again, the inner 68\% range of the posterior distribution is used as an estimation of the error of the best fit. Given the large number of free parameters, the fit did not converge within a reasonable time for most galaxies even on a multiple-core machine. This is one reason why the MCMC error bars are probably too large for many galaxies. The results of both approaches are shown in Fig. \ref{fig:errors}. We only plot the $n_\mathrm{bar}$ parameter. It is clearly visible that the MCMC error bars are much larger than the errors derived with the bootstrap approach.

For the interpretation of these results it is important to keep in mind that the more components are added to the fit the more degenerate it becomes. Therefore, it is more likely to be trapped in local minima in the $\chi^2$ landscape. The MCMC method strongly differs from the bootstrap technique in the sense that it explores the whole parameter space completely independently from the best fit values. However, because of the degeneracies, in many cases the fit did not converge within a reasonable amount of steps. On the contrary, the bootstrap method with Levenberg-Marquardt minimisation starting at the best fit values is apparently not very likely to get out of a local minimum. The main question that we should probably ask ourselves is  what kind of error we want to estimate: uncertainties from observational data, uncertainties from a mismatch between model and reality, or uncertainties from human-made choices for certain model components and initial parameter values. In this context, the results from the MCMC approach demonstrate the spread of possible solutions, if as little assumptions as possible are made. On the other side, bootstrap provides the error given a specific model.

The reason why some of the points in the lower left, best fit bootstrap panel are off the one-to-one line is most probably caused by the different minimisation techniques. The best fit results come from using the slower but more robust Nelder-Mead simplex minimisation while bootstrap uses the fast Levenberg-Marquardt technique. The exact explanation as to why the different algorithms lead to these offsets still needs to be explored in more depth given the complexity of the decompositions. We trust our measurements from the best fit result.

\subsection{Modelling a synthetic galaxy image}
\label{apx:synth}

Another way to estimate the accuracy of the photometric decomposition fitting procedure is to create synthetic images of galaxies and subsequently fit these in the exact same manner as was done for the real galaxies. We show as an example the results for one synthetic galaxy that demonstrates how the input parameters can be retrieved satisfactorily.

The image was created using the \textls{\sc makeimage} module of \textls{\sc imfit}. It takes the same component functions that can be used to fit an image and generates a new image. A psf image can be provided to convolve with the object image before the final output \citep[for further details see][]{Erwin2015a}. We selected six input components: a point source, a bulge, three bar components, and a disc. Additionally, we added a flat sky background. The input values for the model parameters were generated randomly within certain limitations to ensure a realistic galaxy image (for example relative sizes and luminosities). After generating the image with \textls{\sc makeimage} we added on top Poisson noise using the \textls{\sc mknoise} function of \textls{\sc iraf}.

\begin{figure*}
        \centering
        \includegraphics[width=17cm]{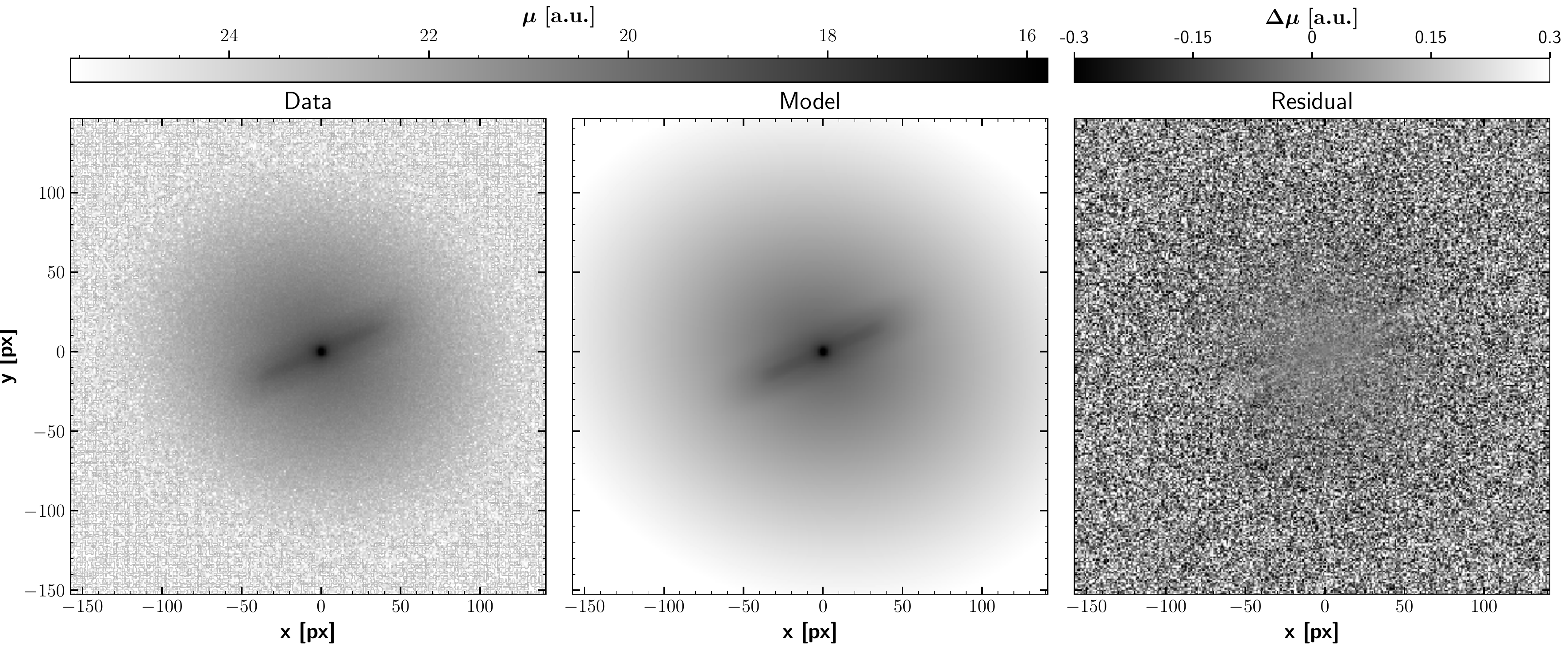}
        \caption{Photometric decomposition of generated artificial galaxy image. From \textit{left} to \textit{right}: data image, model, and residual=data-model. The greyscale mapping in the residual image is stretched in order to show faint details.}
        \label{fig:synth_gal}
\end{figure*}

With no a priori knowledge of the input values, we started the fitting procedure exactly like we did for the real galaxies with a simple point source + exponential disc model. After inspecting the residual images we added other components to the next fit if we had clear visual indications of their presence. As last step we performed bootstrapping with 200 iterations to estimate error intervals. In Table \ref{tbl:art_galaxy} we summarise the results. In Fig. \ref{fig:synth_gal} we show the image of the generated artificial galaxy, the best fitting model and the residual image.

We note that most parameters were successfully retrieved within a range of $\sim 5\%$ deviation including specifically the main bar component \emph{bar1} and its S\'ersic index. Only the bulge and the thick bar component \emph{bar2} show larger deviations from the true values. This can be explained by similar sizes and luminosities of both components and thus degeneracies between the corresponding parameters. We point out, however, that the uncertainties in the photometric fits so derived should be considered as lower limits. This is because the structural components of the synthetic images follow the exact same models employed by the fitting software. This assumption might not necessarily hold in real galaxies.

\begin{table*}
\caption{Synthetic galaxy: Input and best fit values for all model parameters.}
\label{tbl:art_galaxy}
\centering

\begin{tabular}{l c c c c c c}
\hline\hline

Parameter       & Model  &       Input value     &       Best fit value  &        Error   &       Rel. dev.       &       Within errors \\
        & component      &               &               &               &        from input      &       1=true, 0=false \\
(1)      &      (2)     &        (3)     &       (4)     &       (5)     &        (6)     &       (7) \\
\hline
X0       &      Position        &        158.598         &       158.597         &        0.003   &       0.00\%  &       1 \\
Y0       &      &        153.462         &       153.462         &       0.001    &       0.00\%  &       1 \\
\hline
PA [$\degr$]     &       Point source    &       0       &       fixed   &        fixed   &       fixed   &       fixed \\
ell      &       &       0       &       fixed   &       fixed   &       fixed    &       fixed \\
$\mathrm{I_0}$ [counts]  &       &       167000  &       169900  &       4985     &       1.74\%  &       1 \\
$\sigma$         &       &       0.1     &       fixed   &       fixed   &        fixed   &       fixed \\
\hline
PA [$\degr$]     &       Bulge  &        70.970  &       85.154  &       52.559   &       19.99\%         &       1 \\
ell      &       &       0.031   &       0.235   &       0.284   &       646.91\%         &       1 \\
n        &       &       1.800   &       1.646   &       2.240   &       8.53\%   &       1 \\
$\mathrm{I_e}$ [counts]  &       &       17.140  &       16.942  &       21.102   &       1.16\%  &       1 \\
$\mathrm{r_e}$ [px]      &       &       3.200   &       3.848   &       4.295    &       20.26\%         &       1 \\
\hline
PA [$\degr$]     &       Bar1   &        113.450         &       113.489         &        0.088   &       0.03\%  &       1 \\
ell      &       &       0.730   &       0.731   &       0.002   &       0.19\%   &       1 \\
c0       &       &       2.000   &       1.957   &       0.134   &       2.16\%   &       1 \\
n        &       &       0.320   &       0.314   &       0.007   &       1.92\%   &       1 \\
$\mathrm{I_e}$ [counts]  &       &       5.200   &       5.163   &       0.073    &       0.71\%  &       1 \\
$\mathrm{r_e}$ [px]      &       &       32.170  &       32.321  &       0.166    &       0.47\%  &       1 \\
\hline
PA [$\degr$]     &       Bar2   &        113.450         &       113.451         &        2.112   &       0.00\%  &       1 \\
ell      &       &       0.160   &       0.125   &       0.036   &       21.94\%  &       1 \\
c0       &       &       24.200  &       30.707  &       66.273  &       26.89\%  &       1 \\
n        &       &       0.500   &       0.794   &       0.191   &       58.85\%  &       0 \\
$\mathrm{I_e}$ [counts]  &       &       9.200   &       14.577  &       4.075    &       58.45\%         &       0 \\
$\mathrm{r_e}$ [px]      &       &       5.100   &       4.776   &       0.486    &       6.36\%  &       1 \\
\hline
PA [$\degr$]     &       Bar3   &        113.450         &       113.570         &        0.070   &       0.11\%  &       0 \\
ell      &       &       0.900   &       0.898   &       0.002   &       0.22\%   &       1 \\
c0       &       &       1.240   &       1.301   &       0.255   &       4.96\%   &       1 \\
n        &       &       0.160   &       0.154   &       0.010   &       3.87\%   &       1 \\
$\mathrm{I_e}$ [counts]  &       &       7.250   &       7.234   &       0.125    &       0.23\%  &       1 \\
$\mathrm{r_e}$ [px]      &       &       27.530  &       27.409  &       0.253    &       0.44\%  &       1 \\
\hline
PA [$\degr$]     &       Disc   &        70.970  &       71.316  &       0.287    &       0.49\%  &       0 \\
ell      &       &       0.140   &       0.139   &       0.001   &       0.41\%   &       1 \\
$\mathrm{I_0}$ [counts]  &       &       20.000  &       20.021  &       0.073    &       0.11\%  &       1 \\
h [px]   &       &       28.900  &       28.774  &       0.050   &       0.43\%   &       0 \\
\hline
$\mathrm{I_{sky}}$ [counts]      &       Background     &        0.0010  &        0.0004  &       0.0006  &       59.92\%         &       1 \\

\hline                                   
\end{tabular}
\tablefoot{The synthetic galaxy was created with six different structural components and a flat sky background. Poisson noise was added afterwards, but before the fitting procedure. This table gives an overview of the input values as well as all 31 simultaneously fitted free parameters and 3 fixed parameters. (1) Parameter; (2) structural component of the model; (3) input value to create the synthetic galaxy; (4) best fit value; (5) 1$\sigma$ error estimate from bootstrapping with 200 iterations; (6) relative deviation of the best fit value from the input value; and (7) check whether the input value lies within the 1$\sigma$ interval of the best fit result.}
\end{table*}

\clearpage

\section{Major and minor axis surface brightness profiles}

\begin{figure*}[!h]
        \centering
        \includegraphics[width=17cm]{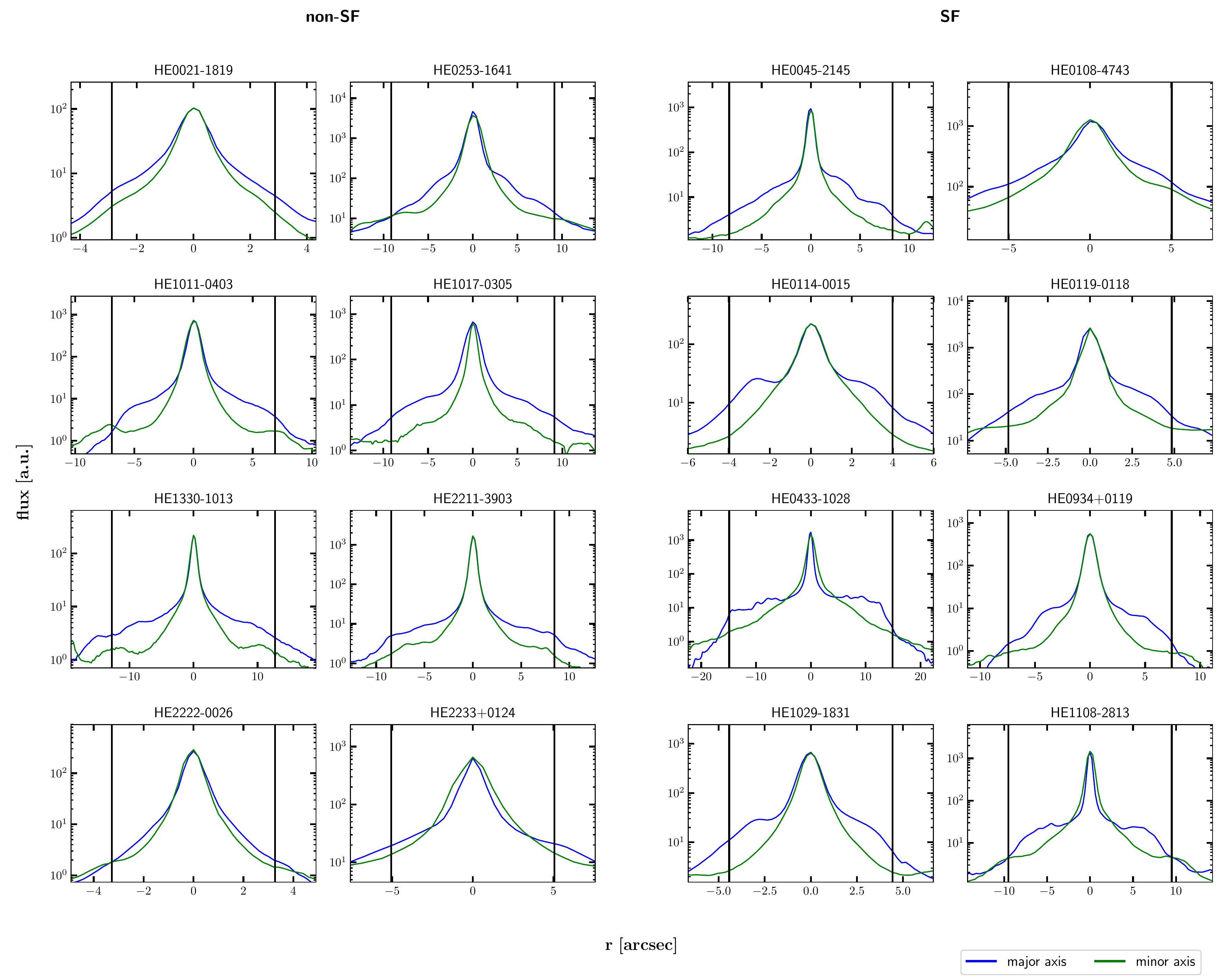}
        \caption{Radial surface brightness profiles along the major (blue) and minor (green) axes of the bar and continued into the disc. The distance on the x-axis is deprojected and scaled to the galaxy plane. The length of the bar is denoted by vertical black lines}
        \label{fig:majorminor}
\end{figure*}

In Fig. \ref{fig:majorminor} we present surface brightness profiles along the major and minor axis of the stellar bars. These were directly extracted from the collapsed $i$-band images of the MUSE cubes. We used these profiles to mimic an analysis similar to \citet{Elmegreen1985} and \citet{Elmegreen1996} and compare our results with theirs; see Sect. \ref{Sect:elmegreen} for further details.

\clearpage

\section{Photometric image decomposition}
\label{apx:2ddecomp}

\begin{figure*}[!h]
        \centering
        \includegraphics[width=17cm]{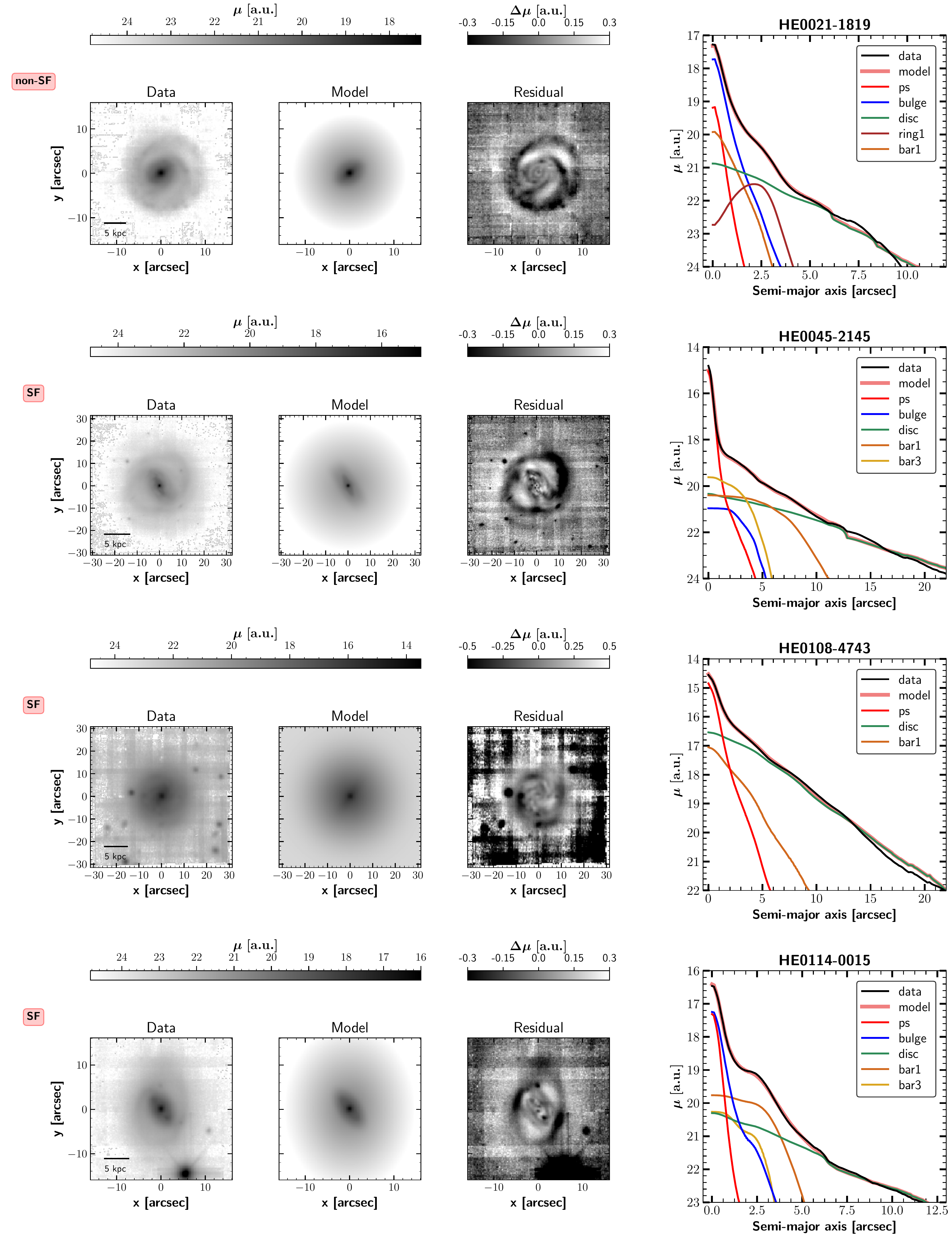}
        \caption{Same as Fig. \ref{fig:2ddecomp}. Photometric image decomposition of the complete sample.}
        \label{fig:2ddecomp1}
\end{figure*}

\begin{figure*}
        \centering
        \includegraphics[width=17cm]{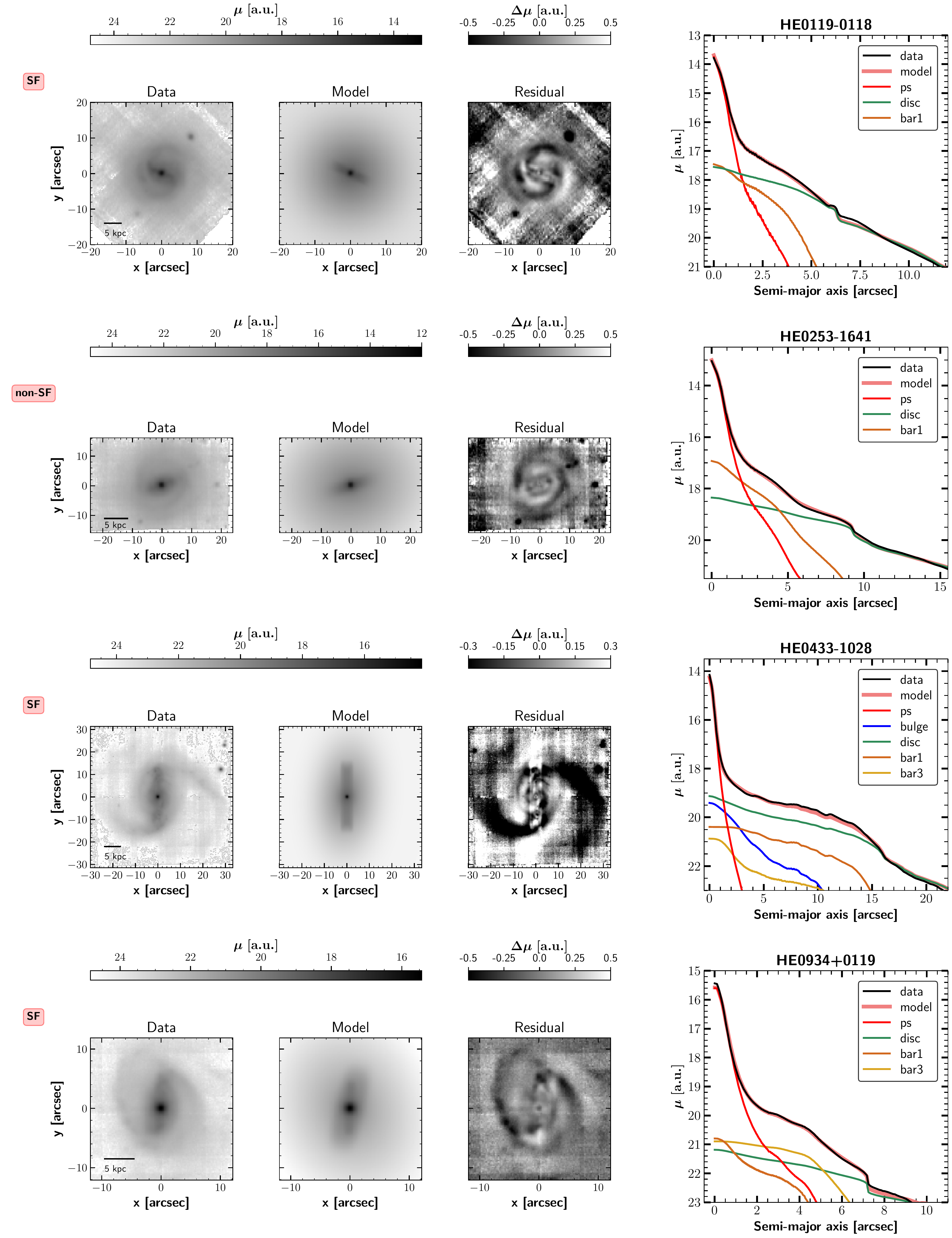}
        \caption{continued.}
        \label{fig:2ddecomp2}
\end{figure*}

\begin{figure*}
        \centering
        \includegraphics[width=17cm]{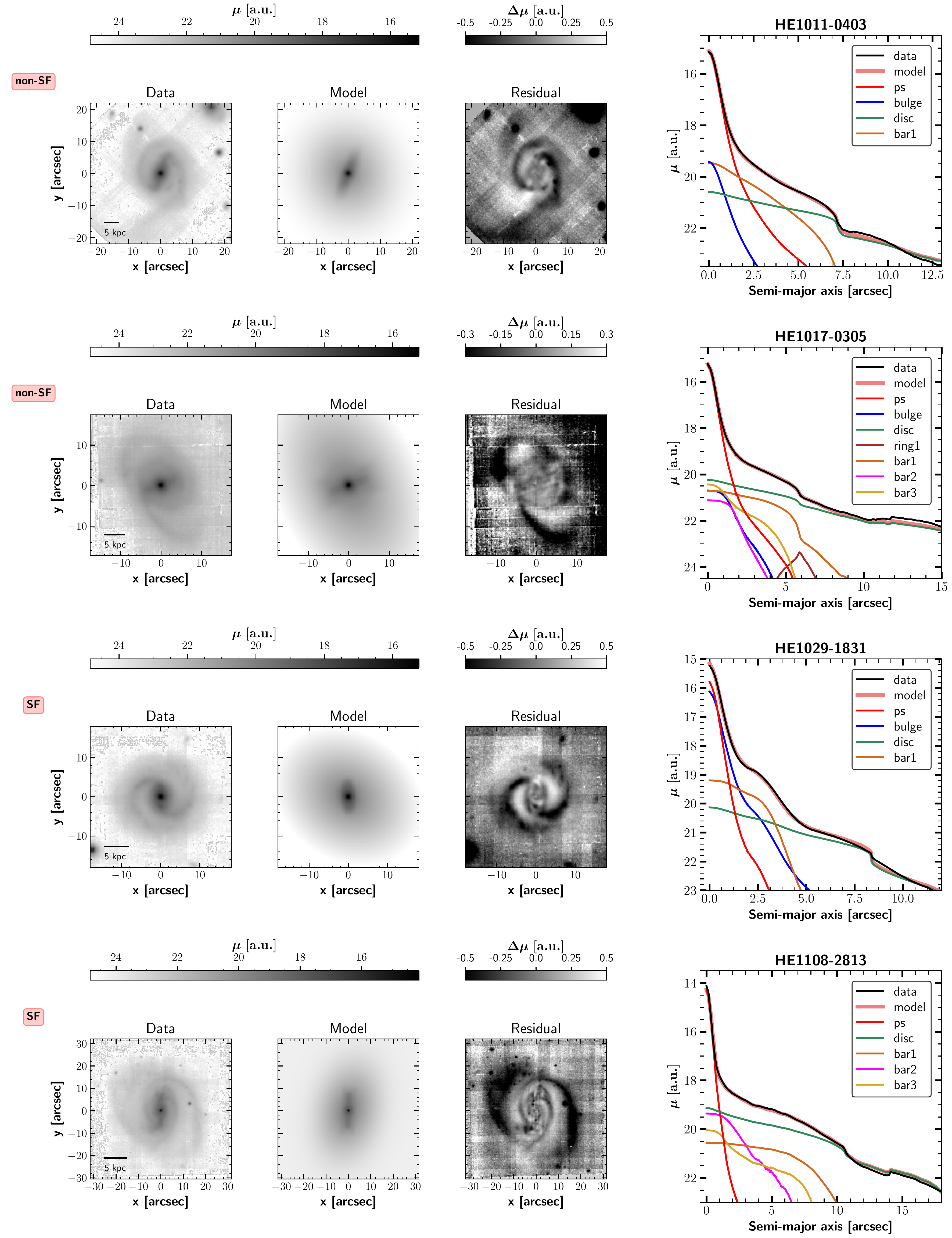}
        \caption{continued.}
        \label{fig:2ddecomp3}
\end{figure*}

\begin{figure*}
        \centering
        \includegraphics[width=17cm]{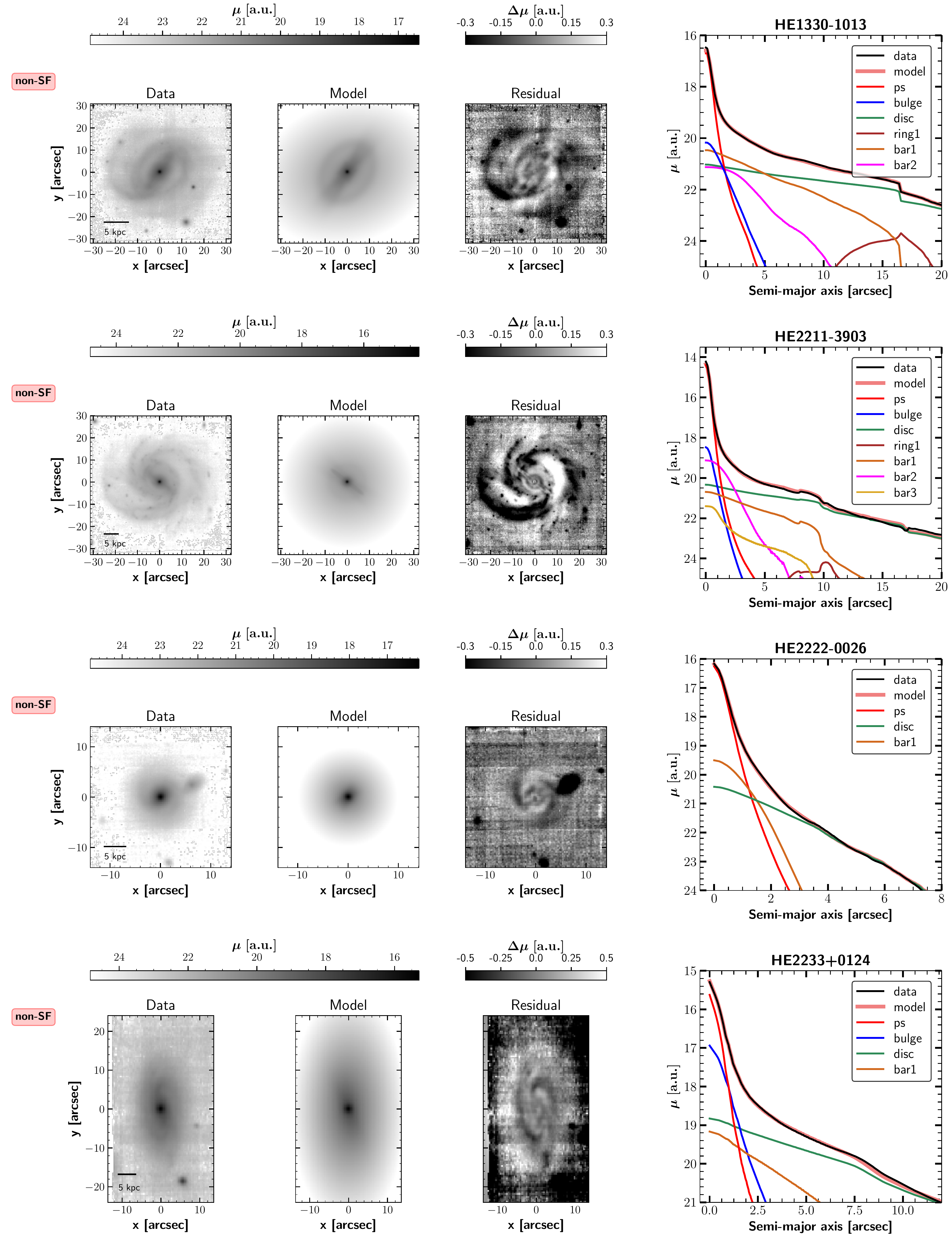}
        \caption{continued.}
        \label{fig:2ddecomp4}
\end{figure*}

\clearpage

\section{Spatial distribution of star formation}
\label{apx:sfr}

In this section we show the spatial distribution of star formation for the entire sample used in this work. The figures are of the same format as Fig. \ref{fig:SFR_map}. The coloured pixels are showing the SFR per spaxel in the MUSE cube overplotted with a $3 \times 9$ grid outlining the region of the bar. This grid is used in the next step of the analysis to rebin the spectra in each cell and estimate the corresponding SFR per bin as described in Sect. \ref{Sect:SFR}. Additionally to Fig. \ref{fig:SFR_map}, here, we also show in blue isophotal contours from the $i$-band image of the galaxy. In the case of galaxy HE1108-2813, the PA of the grid has been adjusted carefully (by $7 \deg$ counterclockwise) to capture fully the star formation that clearly comes from the bar and avoid as much as possible contaminant star formation from spiral arms. This is caused by the offset of the PA of H$\alpha$ as compared to the stellar bar towards the leading edge as discussed in Sect. \ref{Sect:results_SFdist}. The spiral-like pattern seen in many SFR maps follows the typical pattern of infalling gas along the \emph{x1} and \emph{x2} orbits of the bar.

\begin{figure*}[!h]
        \centering
        \includegraphics[width=15cm]{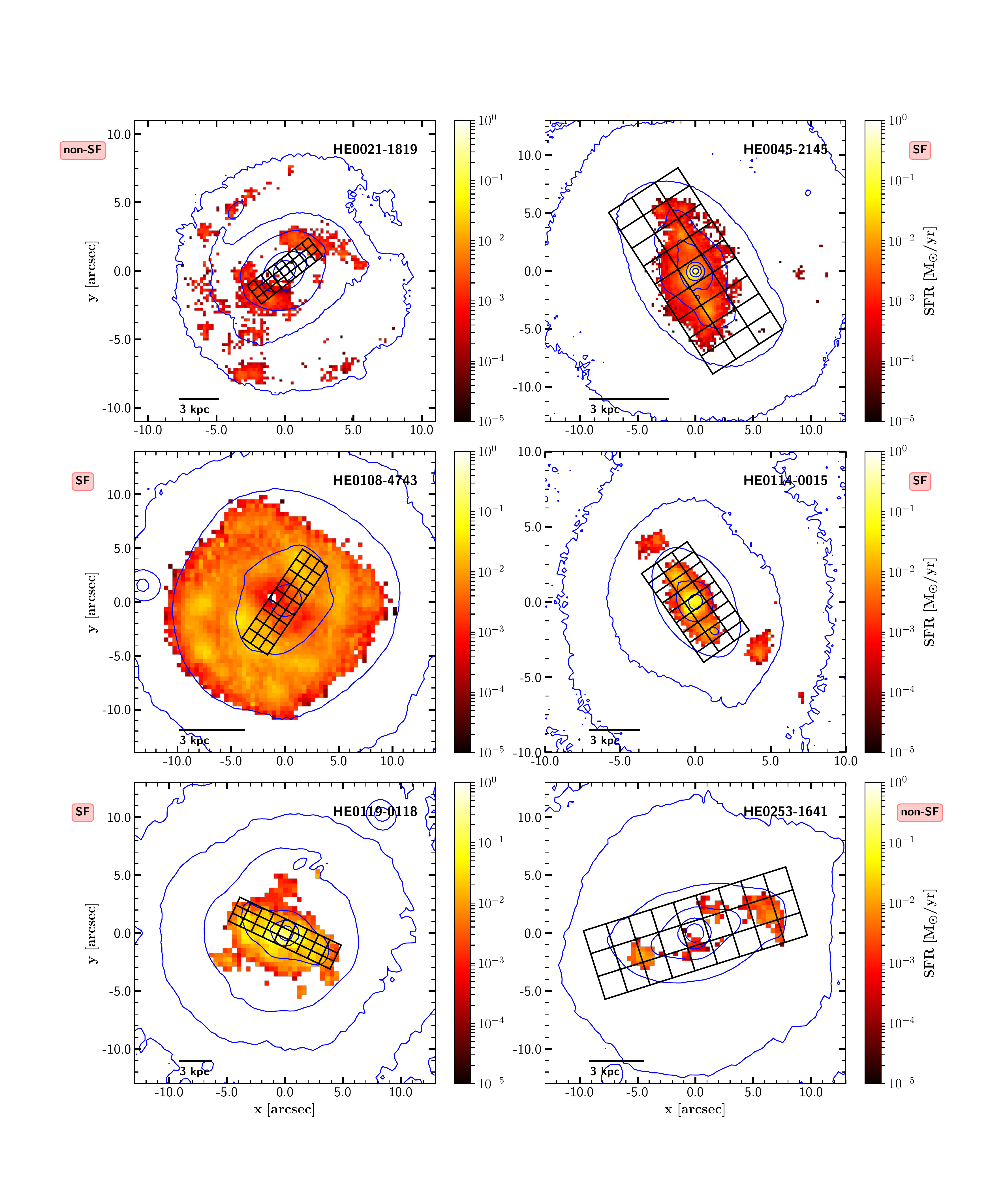}
        \caption{Same as Fig. \ref{fig:SFR_map}. Spatial distribution of SFR of the complete sample. We also show blue contours of the $i$-band image of the galaxy.}
        \label{fig:sfr_map1}
\end{figure*}

\begin{figure*}
        \centering
        \includegraphics[width=15cm]{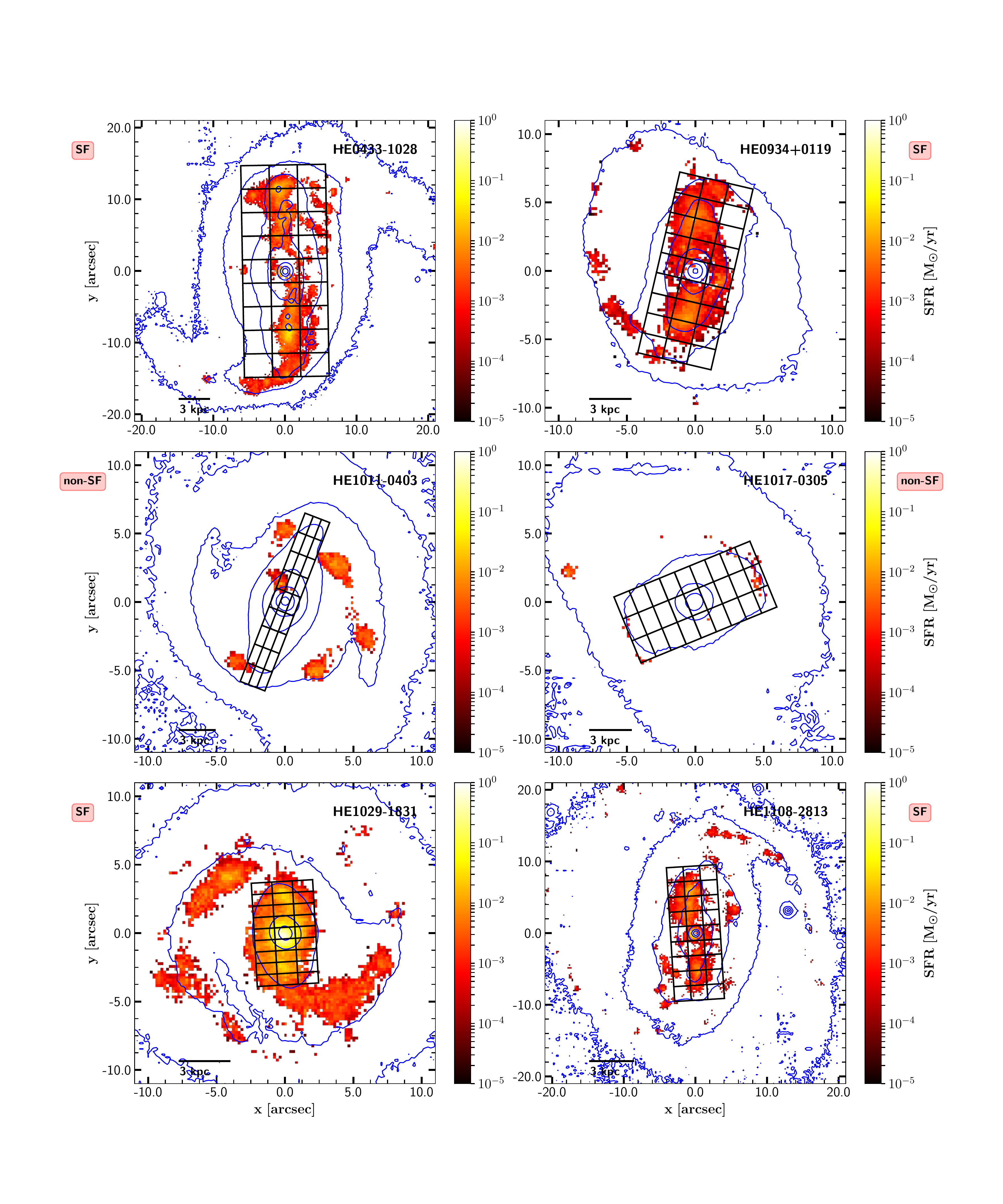}
        \caption{continued.}
        \label{fig:sfr_map2}
\end{figure*}

\begin{figure*}
        \centering
        \includegraphics[width=15cm]{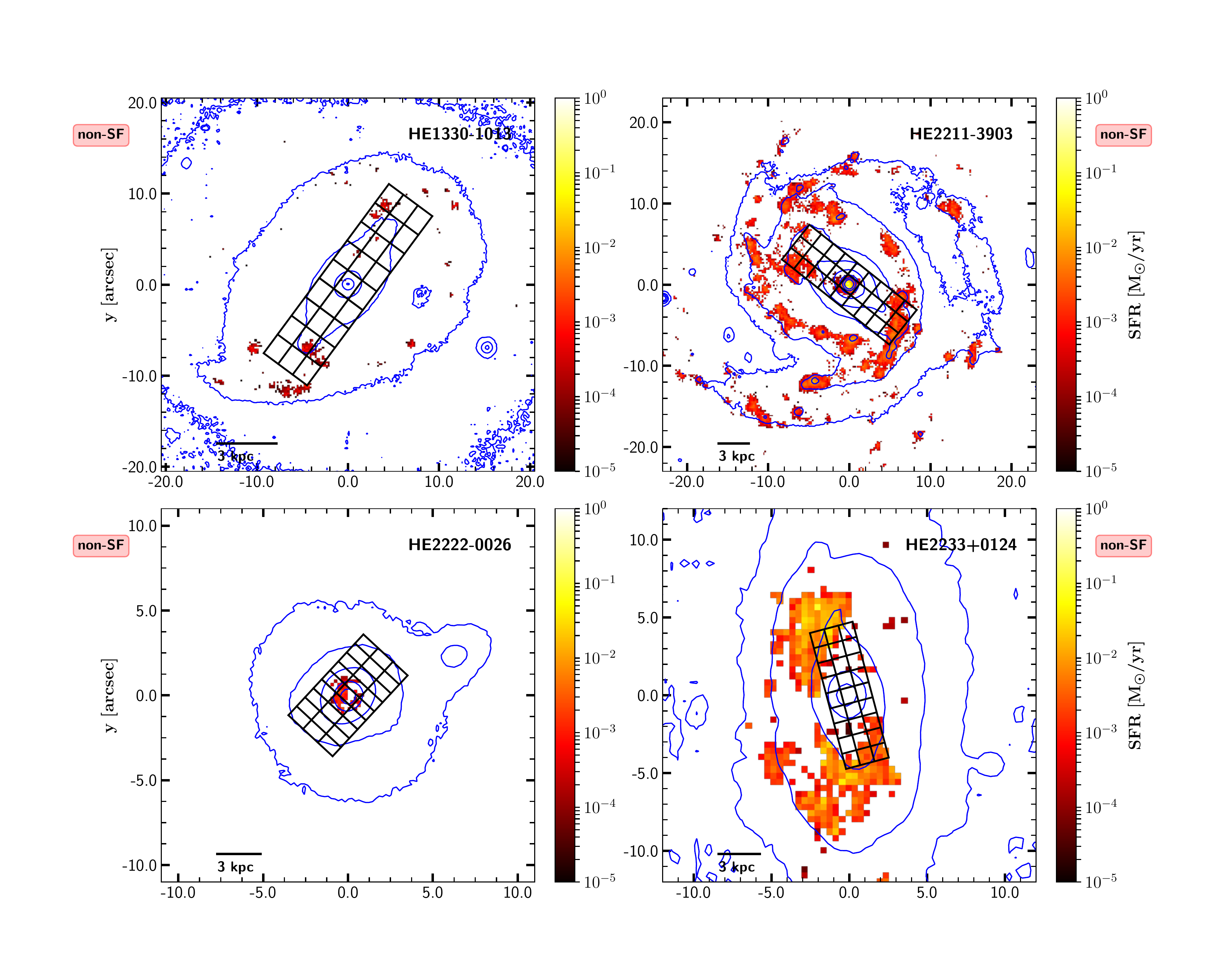}
        \caption{continued.}
        \label{fig:sfr_map3}
\end{figure*}

\end{appendix}

\end{document}